\documentclass[journal]{IEEEtran}

\usepackage{subfigure}
\usepackage{cite}

\ifCLASSINFOpdf
   \usepackage[pdftex]{graphicx}

\else

\fi

\usepackage[cmex10]{amsmath}
\usepackage{url}

\usepackage{amssymb}

\usepackage{colortbl}

\usepackage{epstopdf}

\begin{document}

\title{Coordination-based Medium Access Control with Space-reservation for Wireless Ad Hoc Networks}
\author{Kamal~Rahimi Malekshan,~\IEEEmembership{Student member,~IEEE,}
        Weihua~Zhuang,~\IEEEmembership{Fellow,~IEEE}\\
        and~Yves~Lostanlen,~\IEEEmembership{Senior member,~IEEE}
\thanks{
This work was supported by a research grant from the Natural Sciences and Engineering Research Council (NSERC) of Canada.

K. Rahimi Malekshan and W. Zhuang are with the Department of Electrical and Computer Engineering,
University of Waterloo, Canada (e-mail: \{krahimim, wzhuang\}@uwaterloo.ca). Y. Lostanlen is with SIRADEL North America, Toronto, Canada (e-mail: yves.lostanlen@ieee.org).

}
}
\maketitle

\begin{abstract}
Efficient radio spectrum utilization and low energy consumption in mobile devices are essential in developing next generation wireless networks. This paper presents a new medium access control (MAC) mechanism to enhance spectrum efficiency and reduce energy consumption in a wireless ad hoc network. A set of coordinator nodes, distributed in the network area, periodically schedule contention-free time slots for all data transmissions/receptions in the network, based on transmission requests from source nodes. Adjacent coordinators exchange scheduling information to effectively increase spatial spectrum reuse and avoid transmission collisions. Moreover, the proposed MAC scheme allows a node to put its radio interface into a sleep mode when it is not transmitting/receiving a packet, in order to reduce energy consumption. Simulation results demonstrate that the proposed scheme achieves substantially higher throughput and has significantly lower energy consumption in comparison with existing schemes.
\end{abstract}

\begin{IEEEkeywords}
Medium access control, wireless ad hoc networks, spectrum efficiency, energy efficiency.
\end{IEEEkeywords}

\IEEEpeerreviewmaketitle

\section{Introduction}

The number of mobile devices and the volume of mobile data traffic have been constantly increasing. It is forecasted that there will be over 10 billion interconnected mobile devices, including machine-to-machine (M2M) modules, by 2018 \cite{VNI}. Overall mobile data traffic is expected to grow nearly 11-fold by 2018 from that in 2013 \cite{VNI}. To meet the increasing growth of mobile data traffic, it is essential to efficiently utilize network resources in the next generation wireless networks. A short communication range in small cells (or WiFi) for hot-spot mobile communications is a key to increase network capacity via spatial spectrum reuse. Such a dense network of mobile nodes and access points and emerging M2M communications necessitate establishing self-organizing ad hoc networks to opportunistically leverage spectrum. Yet, energy consumption by radio interfaces should be minimized, because of limited battery storage of mobile devices. In \cite{Kamal2}, we present a new energy efficient MAC protocol with high throughput and low packet transmission delay for a fully connected network, in which only one node can transmit at each time instance over the radio channel.

In a wireless ad hoc network, nodes that are not in the communication range of each other cannot hear each others' transmissions. However, their transmission may interfere each other at the receiver nodes. On the other hand, nodes that are sufficiently far apart in space can transmit simultaneously without a collision (i.e., spatial spectrum reuse is possible). Thus, an effective medium access control (MAC) scheme for a wireless network should have the following features:
\begin{enumerate}
  \item It should prevent simultaneous transmission of interfering links. Otherwise, one or more of the transmissions will fail because of transmission collision, which results in wastage of radio bandwidth and energy;\label{item1}
  \item It should allow simultaneous transmissions of non-interfering links for spatial reuse of the radio channel, because preventing non-interfering links from simultaneous transmission will unnecessarily degrade throughput of the network.\label{item2}
\end{enumerate}
When a MAC scheme fails to accomplish the first feature, the \begin{it}hidden terminal\end{it} problem arises. On the other hand, when a MAC scheme does not have the second feature, the \begin{it}exposed terminal\end{it} problem occurs. A TDMA (Time Division Multiple Access) MAC scheme can potentially solve both the \begin{it}hidden terminal\end{it} and \begin{it}exposed terminal\end{it} problems in a wireless ad hoc network. However, finding an efficient time schedule requires a central controller and the optimal solution is NP-hard \cite{RAND,NP-hard}. Moreover, in a wireless ad hoc network, the traffic load and network topology change with time, which makes the static TDMA very inefficient. In addition, reassignment of channel time imposes a large overhead and requires global changes. The CSMA (Carrier Sense Multiple Access) MAC is commonly used in wireless ad hoc (and wireless local area) networks because of its flexibility and simplicity. However, it suffers from transmission collision and contention overhead, and cannot resolve the hidden and exposed terminal problems in a wireless ad hoc network. The hidden terminal problem can be avoided by increasing the carrier sensing range \cite{Solve-HN-CS}, which however aggravates the exposed terminal problem and results in wastage of radio bandwidth. The RTS/CTS (Request-to-send/clear-to-send) mechanism is used in \cite{Standard,GMAC,S-MAC,T-MAC,EN-Jiang} to mitigate the hidden terminal problem. However, this mechanism imposes a significant amount of overhead in bandwidth and energy.

Radio interface is a main source of energy consumption in mobile devices, which can quickly drain the device's limited battery \cite{Tsao:2011:, Greening, Pering:2006:CRP:1134680.1134704,WiZi-Cloud}. For instance, the WiFi radio consumes more than 70\% of total energy in a smartphone when the screen is off \cite{Pering:2006:CRP:1134680.1134704}, which is reduced to 44.5\% in the power saving mode. A radio interface can be in one of the following modes: transmit, receive, idle, and sleep. A radio interface consumes a significant amount of energy in the idle mode, in which it is neither transmitting nor receiving a packet. For instance, a \begin{it}Cisco Aironet 350 series\end{it} WLAN adapter \cite{Cisco} consumes 2.25W, 1.25W, 1.25W, and 0.075W in the transmit, receive, idle, and sleep modes respectively. In order to reduce energy consumption, nodes should periodically put their radio interfaces into the sleep mode. While a radio interface is in the sleep mode, the node cannot receive incoming packets. Thus, a transmitter node should be aware of the receiver node's status to successfully deliver a packet.

In a cellular network, network area is partitioned into cells and nodes inside a cell only communicate with the cell base station (BS) at the cell center. The BS schedules all transmissions/receptions to and from nodes (downlink and uplink) inside its cell. Therefore, transmission collisions are prevented among nodes in the cell and idle listening energy consumption of mobile nodes is minimized, because of deterministic transmission/reception time which is assigned by the BS. In the conventional cellular networks, each cell is assigned a fraction of total available radio spectrum to avoid inter-cell interference. For instance, in GSM a cell commonly uses one-fourth of total available radio spectrum (frequency reuse factor 4) to prevent inter-cell interference. Several inter-cell interference coordination techniques are proposed to improve network performance of cellular systems using fractional frequency reuse \cite{Hamza,3GPP}. In fractional frequency reuse, the total available radio spectrum is used for transmissions to and from the nodes close to the BS at the central region of a cell, but a fraction of spectrum is used for transmissions to and from nodes that are outside the central region of the cell, in order to reduce inter-cell interference \cite{Hamza,3GPP,Ismail,Ran}. The dense deployment of small cells in the next generation of wireless networks and the direct device-to-device (D2D) and M2M communications form communication links in an ad hoc manner, which require a new MAC mechanism to efficiently utilize the shared radio spectrum and minimize power consumption.

In this paper, we propose a novel medium access mechanism for a wireless ad hoc network with arbitrary communication links. Table \ref{Table3} compares main characteristics of the proposed MAC mechanism and existing approaches. The proposed scheme combines the deterministic transmission/reception feature of cellular networks and the opportunistic spectrum access feature of WiFi networks to efficiently utilize shared spectrum and minimize energy consumption. A set of coordinators distributed in the network area are chosen to dynamically coordinate contention-free time slots for all data transmissions/receptions based on transmission requests from source nodes. Each coordinator periodically broadcasts a scheduling packet to schedule all transmissions/receptions in its proximity. For each scheduled transmission/reception, the space around the receiver node is reserved to avoid transmission collision and enhance spatial radio spectrum reuse. A coordinator collects nodes' transmission requests and overhears the scheduling packets of its neighboring coordinators. Accordingly, each coordinator schedules a transmission/reception only if the transmission of the source node does not interfere with other scheduled receptions and the other scheduled transmissions do not interfere with the reception at the destination. Dynamic assignment of the shared radio spectrum and adequate spatial spectrum reuse increase spectrum efficiency. Moreover, a deterministic transmission/reception time warrants nodes to put their radio interface into the sleep mode when they are neither transmitting nor receiving a packet, which reduces energy consumption. Comparing with existing schemes, the proposed MAC provides significantly higher throughput and greatly reduces node energy consumption.

The rest of this paper is organized as follows: Section II reviews related works. The system model is presented in Section III. In Section IV, we describe the proposed MAC mechanism. Simulation results are presented in Section V to evaluate the performance of proposed MAC solution in comparison with existing schemes. Finally, Section VI concludes this research.

\begin{table}[tb]
    \begin{center}
    \caption{Comparison of proposed MAC mechanism with existing approaches}
     \label{Table3}
        \includegraphics[width=3.54in, trim=2.1cm 19cm 1.9cm 2.5cm, clip=true]{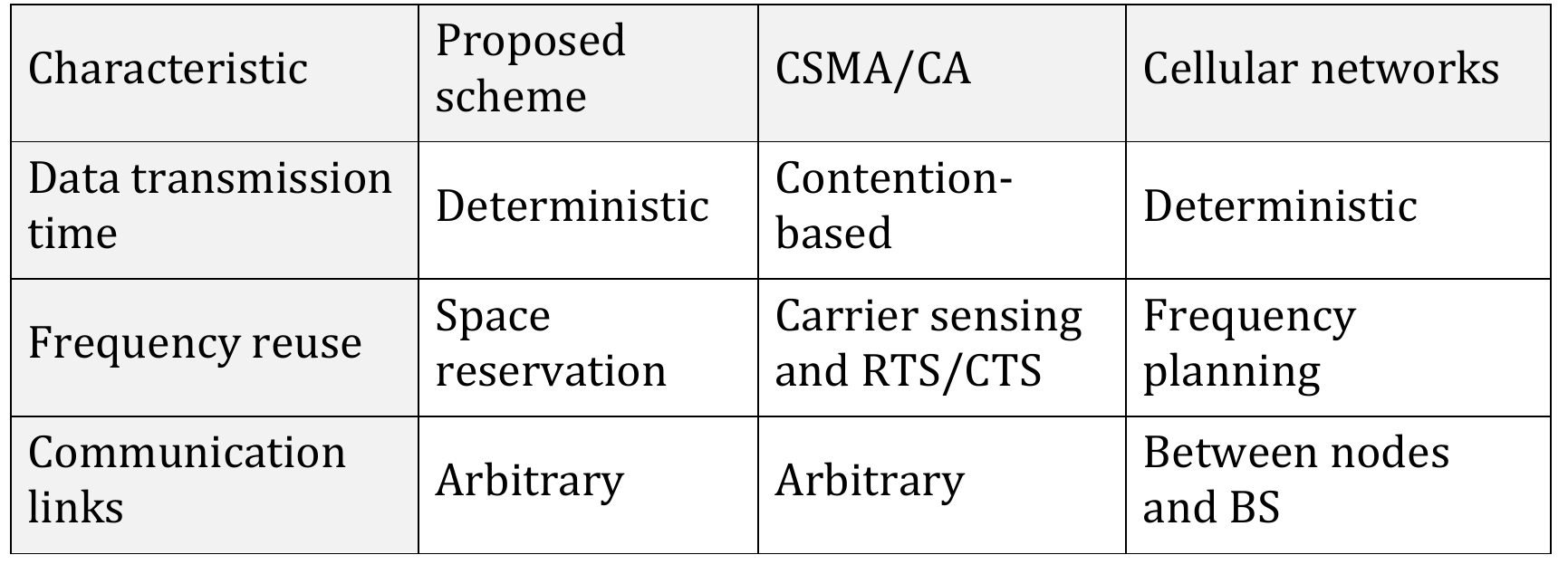}
    \end{center}
\end{table}

\section{Related Works}
A dynamic TDMA MAC scheme is proposed in \cite{ADHOC-MAC,Hassan}. Time is partitioned into frames of $F$ slots. Every node acquires a transmission slot in each frame, in which it transmits a packet to inform the other nodes of the time slots that it will transmit/receive data packets (frame information). A node can reserve additional transmission slots using ALOHA and/or via broadcasting its frame information. A node can reserve a new time slot only if none of the neighboring nodes has announced a transmission/reception in that time slot in the previous frame. This mechanism can mitigate the hidden terminal problem; however the imposed overhead of transmitting frame information by every node in each frame reduces network throughput and increases energy consumption. A hybrid TDMA-CSMA MAC scheme is proposed in \cite{Z-MAC} using CSMA as the baseline MAC scheme. A transmission time slot is assigned to each node such that none of the interfering nodes are assigned a same transmission slot. At each time slot, the owner has a higher priority to transmit a packet. If a node experiences successive collisions because of hidden nodes, it will transmit a request packet to prevent the interfering nodes from transmission in its assigned transmission slot for a requested period of time.

The existing single-channel energy saving MAC protocols for a wireless ad hoc network can be classified into synchronous and asynchronous energy saving protocols. In the synchronous energy saving schemes \cite{Standard, DPSM, TMMAC}, all nodes are synchronized in time and time is partitioned into beacon intervals. All nodes wake up simultaneously at the beginning of each beacon interval, in an ATIM (Ad hoc traffic indication message) window, to exchange ATIM packets which are transmitted by the sender nodes to inform their intended receivers of buffered packets. During the rest of beacon interval (i.e., the communication period), the nodes that have packets to transmit/receive stay awake to communicate. Other nodes switch their radio interfaces into the sleep mode to save energy. In the asynchronous energy saving schemes \cite{Tseng, QuorumAsynchronized}, nodes are not synchronized in time and each node has its own clock. Each node evenly divides its time into beacon intervals. There are two types of beacon intervals: active beacon intervals and energy saving beacon intervals. An active beacon interval starts with a beacon window, during which the node should contend to transmit a beacon including its clock and wake up pattern. Followed by the beacon window is an MTIM (Multi-hop traffic indication message) window (similar to the ATIM window in the synchronous schemes), in which the nodes with buffered packets notify the intended receivers. After that, each node stays awake for the rest of beacon interval to receive the beacons transmitted by other nodes. An energy saving beacon interval starts by an MTIM window, and after the MTIM window, the node can power off if it has no packet to send or receive. The patterns of active and energy saving beacon intervals should be chosen to ensure that the beacon of each node will be heard by every single-hop node at least once in a predefined period of time.

Existing energy saving schemes for wireless ad hoc networks require all the nodes to stay awake during the ATIM (or MTIM) window in each beacon interval, and every node with a packet for transmission has to contend to send a notification packet to its intended receiver at each beacon interval. The ATIM (or MTIM) window overhead decreases the communication period in each beacon interval and increases energy consumption. Further, the ATIM (or MTIM) size significantly affects the network performance and should be adjusted based on the networking condition. How to choose an optimal ATIM (or MTIM) size is an open issue.

The synchronous energy saving mode requires all nodes to be synchronized in time. Time synchronization in a wireless ad hoc network is challenging because propagation delays are long and the network may temporarily be partitioned. Although asynchronous power saving schemes do not need the synchronization among nodes, no need for synchronization comes at expense of a requirement for periodically active beacon intervals and more beacon transmissions than in synchronous power saving schemes. The more frequent beacon transmissions and periodically active beacon intervals cause more transmission overhead and more energy consumption in asynchronous power saving schemes, in comparison with synchronous power saving schemes.

In the existing power saving MAC schemes, the contention and collision overhead during the communication period degrades the network throughput and increases energy consumption. In a wireless network, the collision rate is further increased because of the \begin{it}hidden terminal\end{it} problem, which further decreases the performance of CSMA/CA MAC used in the existing power saving protocols. Moreover, in the CSMA/CA MAC, the radio channel is not efficiently utilized because of the \begin{it}exposed terminal\end{it} problem, which further reduces the network performance. Even though the TMMAC \cite{TMMAC} uses a contention-free MAC protocol in the communication period to reduce overhead and energy consumption, it cannot fully utilize the radio channel bandwidth because nodes reserve time slots independently without coordination. Also, the TMMAC requires exchanging of three control packets (ATIM/ATIM-ACK/ATIM-RES) between the source and destination nodes in the ATIM window, which degrades channel utilization and increases energy consumption.

In this work, we propose a new MAC mechanism to achieve high throughput and low energy consumption in a wireless ad hoc network. Using a set of coordinators, the proposed MAC scheme dynamically reserves channel in both time and space domains for data transmissions/receptions based on nodes' transmission requests. Exchanging scheduling information among adjacent coordinators empowers the proposed MAC scheme to effectively increase spatial spectrum reuse and prevent transmission collision. Also, periodic assignment of contention-free data transmission/reception time slots enables a node to put its radio interface into the sleep mode when it is not transmitting/receiving a packet, which significantly reduces energy consumption.

\section{System Model}
Consider a wireless ad hoc network with $N$ nodes where all nodes are not in the communication range of each other. We focus on single-channel single-hop transmissions as, at the MAC layer, each node communicates with one or more of its one-hop neighboring nodes. Let $l_{ij}$ denote single-hop link from source node $i\in\{1,2,...,N\}$ to destination node $j\in\{1,2,...,N\}, i\neq j$. We denote the distance between the source and destination nodes of $l_{ij}$ by $d_{ij}$. The channel gain between source node $i$ and destination node $j$ is $h_{ij}=cd_{ij}^{-\alpha}$, where $c$ is a constant and $\alpha$ is the path loss exponent. Let $\bar{\boldsymbol{p}}=(p_1,p_2,...,p_N)$ denote the transmission power vector, where $p_{i}, i\in\{1,2,...,N\}$, denotes the transmission power level of source node $i$. Let $\bar{\boldsymbol{u}}=(u_1,u_2,...,u_N)$ denote the transmission vector, where $u_{i}=1$ denotes that node $i$ is scheduled for transmission and $u_{i}=0$ otherwise. Thus, the signal to noise plus interference ratio (SINR) at the destination of link $l_{ij}$ is given by
\begin{equation}\label{P4-20}
  \gamma_{ij}=\frac{u_i p_{i}h_{ij}}{N_0+\sum_{k\neq i}u_k p_{k}h_{kj}}
\end{equation}
where $N_0$ is background noise power and $\sum_{k\neq i}u_{kj}p_{k}h_{kj}\triangleq I_{ij}$ is the amount of interference at the destination of link $l_{ij}$.

All control/scheduling packets are transmitted at power level $P_s$ at rate $R_s$ bps and all data packets are transmitted at power level $P_d$ at rate $R_d$ bps. The corresponding minimum required SINR at a receiver node to successfully receive control/scheduling and data packets are denoted by $\Gamma_s$ and $\Gamma_d$ respectively.

\section{Medium Access Control}
\begin{figure}[t]
  \centering
  \includegraphics[width=2.4in, trim=5cm 2cm 9cm 4cm, clip=true]{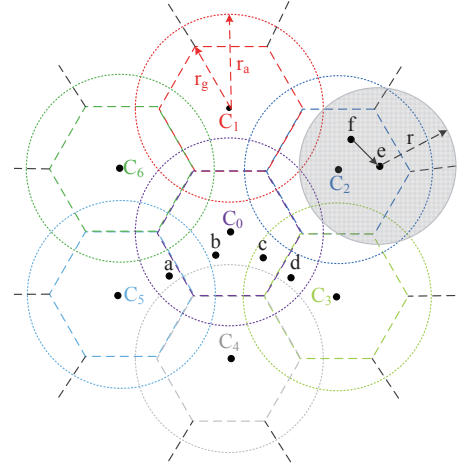}\\
  \caption{Partitioning the network area into hexagonal cells, where $C_i$, $i\in\{0, 1, 2,...,m\}$, denotes the coordinator of cell $i$, the dotted circle centred at $C_{i}$ shows the area that $C_{i}$ broadcasts all scheduled transmissions/receptions, and the shaded area shows the space reserved for transmission from node $f$ to node $e$.}
  \label{PMAC1}
\end{figure}

In order to efficiently utilize the radio channel and minimize energy consumption in a wireless ad hoc network, we use the following main strategies:
\begin{enumerate}
  \item Dynamic coordination of access to the shared medium based on instantaneous traffic load by a set of coordinators distributed in the network area;
  \item Preventing transmission collisions and minimizing idle listening power consumption by periodic assignment of deterministic time slots for data transmissions/receptions;
  \item Effective spatial channel reuse by space-reservation for scheduled transmissions/receptions and by exchanging scheduling information among adjacent coordinators.
\end{enumerate}
The network coverage area is partitioned into hexagonal cells, as shown in Figure \ref{PMAC1}. The distance between the center and a vertex of a cell is denoted by $r_{g}$, which is set such that $r_g \geq \max_{d_{ij}\in \mathcal{L}} d_{ij}$, where $\mathcal{L}$ is the set of single-hop links in the network. Therefore, the source and destination nodes of each single-hop link are either in one cell or adjacent cells. A node at the center of each cell coordinates all the transmissions/receptions for nodes inside the cell. We assume that coordinators have higher energy capacity and do not move frequently (e.g., access points). Thus, the network planning does not need to be updated frequently.

\begin{figure}[t]
  \centering
  \includegraphics[width=3.7in, trim=3.7cm 12cm 3.1cm 2.6cm, clip=true]{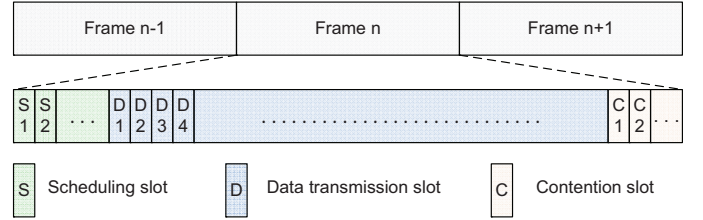}\\
  \caption{Structure of one frame of the proposed scheme.}
  \label{FrameStructure}
\end{figure}

All nodes are synchronized in time, and time is partitioned into frames. Figure \ref{FrameStructure} shows the structure of a frame. Each frame consists of three types of time slots, i.e., \begin{it}scheduling slots\end{it}, \begin{it}contention-free slots\end{it}, and \begin{it}contention slots\end{it}. In scheduling time slots, located at the beginning of each frame, coordinators transmit scheduling packets to coordinate transmissions/receptions of the current frame. The scheduling packet of a coordinator should be received by all nodes in the cell and adjacent coordinators. Data packet transmissions/receptions take place in contention-free time slots, as scheduled by coordinators. A source node scheduled for transmission in contention-free slots can notify the cell coordinator of its transmission request for the next frame by including information in the header of one data packet. During contention slots, source nodes that want to initiate a new transmission contend with each other to send a transmission request to the cell coordinator. In the following, we describe transmission policy in each time slot, and then the detail operation of the MAC protocol.


\begin{figure}
\centering
\subfigure[]{\includegraphics[width=2.4in, trim=5cm 1cm 3cm 1cm, clip=true]{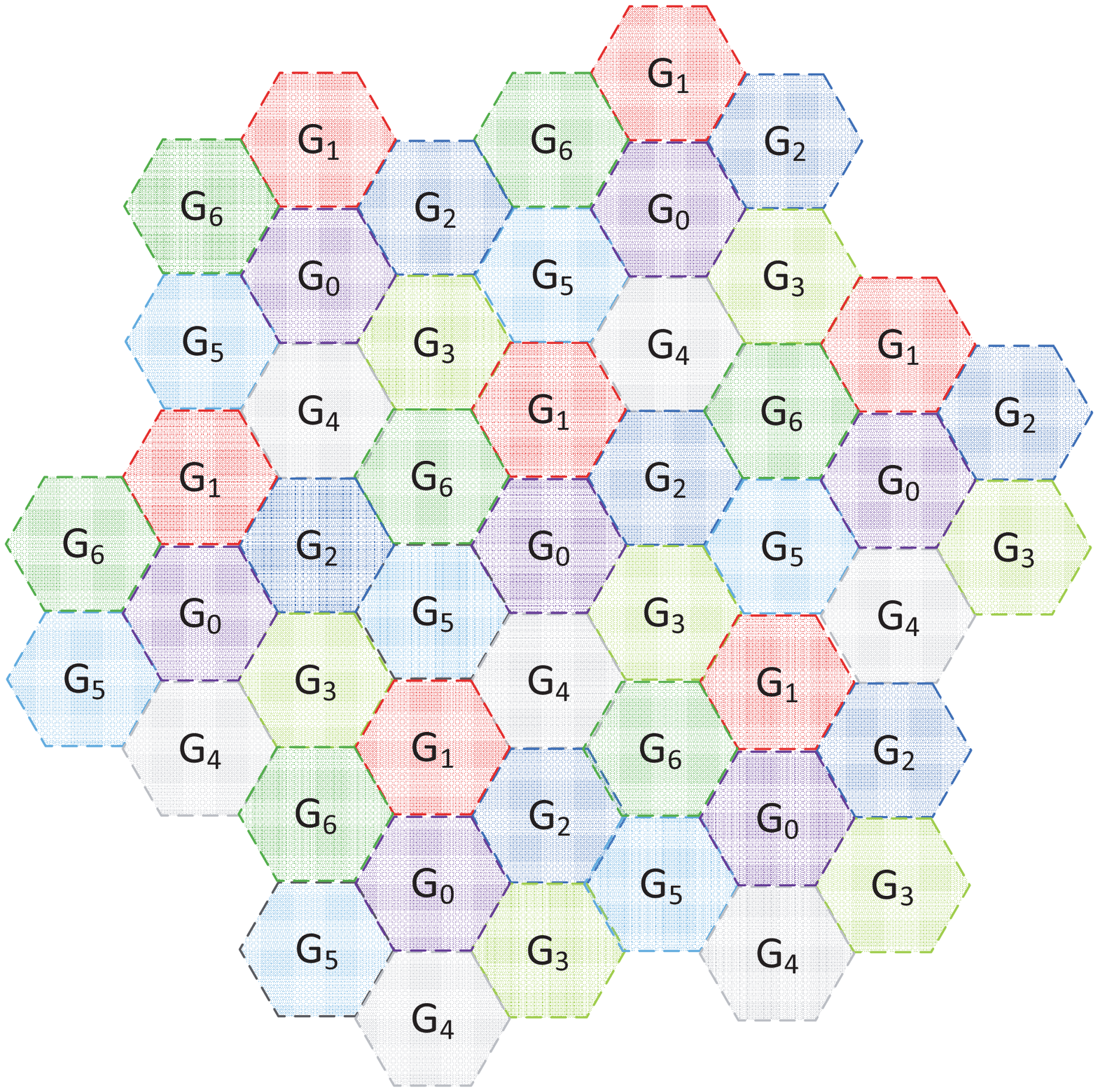}\label{CSLOTS}}
\subfigure[]{\includegraphics[width=3.7in, trim=2.2cm 13.4cm 4.9cm 3.5cm, clip=true]{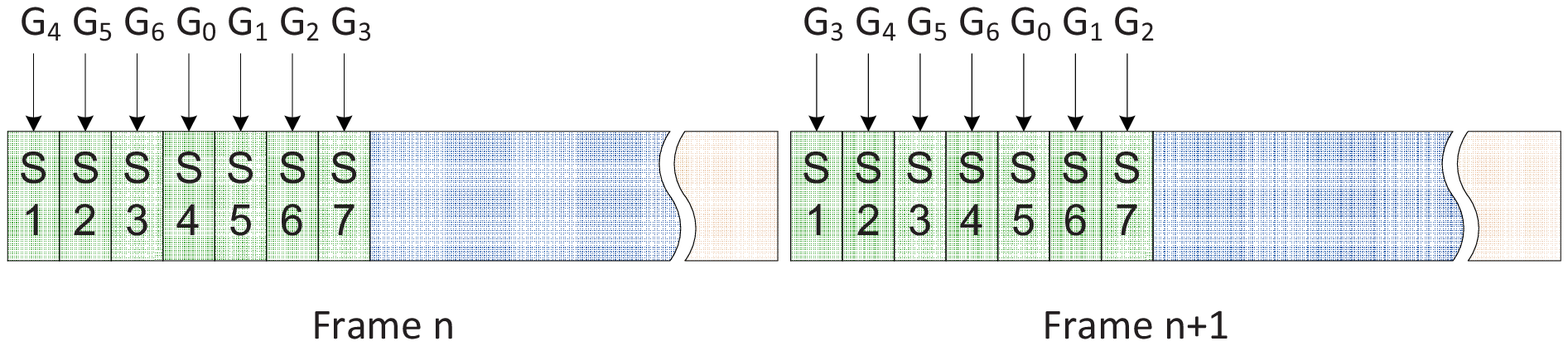}\label{FrameStructureEx}}
\caption{Assignment of scheduling time slots to coordinators, in which a scheduling time slot is assigned to all the coordinators of cells of a same group/color.} \label{ASLOTS}
\end{figure}

\subsection{Transmission policies in the different time slots}
\begin{it}Scheduling slots\end{it}: Scheduling time slots are assigned to coordinators such that a scheduling time slot, assigned to a coordinator, is not assigned to any other two-hop neighboring coordinator. Let $S_j, j\in\{1,2,..,k\}$, denote the $j$th scheduling slot in a frame and $G_i, i\in\{0,2,3,...,k-1\}$, denote the set of coordinators that can be assigned same scheduling time slot. Similar to frequency reuse in cellular networks, with $k$ ($=7$) scheduling time slots, as illustrated in Figure \ref{ASLOTS}, every coordinator can acquire a scheduling time slot that is not assigned to any other two-hop neighboring coordinator node. To ensure fair channel access for nodes in different cells, we change transmission order of coordinators in each frame as illustrated in Figure \ref{FrameStructureEx}. In frame $n$, coordinators $G_i$ are assigned the $j^*$th scheduling time slot ($S_{j^*}$) where $j^*=(n\mod i+1)+1$. Moreover, the size ($r_g$) of cells, transmission power level ($P_s$) for scheduling packets, and data transmission rate ($R_s$) of scheduling packets are selected such that a scheduling packet is received by all nodes inside the cell and all adjacent coordinators (with $\text{SINR}\geq \Gamma_s$).

\begin{it}Contention-free slots\end{it}: Data packet transmissions/receptions are scheduled in contention-free time slots. For each scheduled transmission/reception, no other node should be scheduled for transmission in a reserved area around the receiver to guarantee required SINR at the destination. The shaded area in Figure \ref{PMAC1} shows the reserved space for transmission from node $f$ to node $e$, where no other node is scheduled for transmission in the area to guarantee the required SINR at node $e$. The reserved area for a scheduled link can be parts of several adjacent cells (as in Figure \ref{PMAC1}), which is determined by exchanging real-time scheduling information among adjacent coordinators. The proposed space-reservation mechanism is to provide effective spatial spectrum reuse to improve spectrum efficiency while avoiding transmission collisions. In addition, for each scheduled source node in the current frame, the space around the cell coordinator is reserved during one contention-free slot to enure that the cell coordinator receives the transmission request of source node (for the next frame) that is included in the header of a data packet. When a link is scheduled for transmission, all other nodes in the reserved area around the receiver (and around coordinators) are denoted as interfering nodes and should not be scheduled for transmission. Let $r(d)$ denote the radius of the circular reserved area centered at the receiver node when the distance between the transmitter and receiver is $d$. The amount of interference imposed on the receiver due to transmissions outside the reserved area has an upper bound given by
\begin{equation}\label{78}
  I(d)\leq \hat{I}(d)\triangleq c' \frac{c P_d} {{r(d)}^{\alpha}},
\end{equation}
where $c'$ is a constant and depends on the node density and network traffic load. Therefore, the received SINR at the destination can be represented by
\begin{equation}\label{89}
  \gamma \geq \frac{\frac{cP_d}{{d}^{\alpha}}}{N_0+\hat{I}(d)}.
\end{equation}
Using (\ref{78}) and (\ref{89}), the minimum radius of the reserved circular area centred at the receiver to guarantee $\gamma\geq\Gamma_d$ can be calculated as
\begin{equation}\label{90}
  r(d)=\Big(\frac{c'c P_d}{\frac{c P_d}{{d}^{\alpha}\Gamma_d}-N_0}\Big)^{1/\alpha}.
\end{equation}
Under the assumption $\hat{I}(d)\gg N_0$,
\begin{equation}\label{91}
  r(d)\thickapprox {(c'\Gamma_d)}^{1/\alpha}d.
\end{equation}
According to (\ref{90}) and (\ref{91}), as $c'$ increases, the reserved circular area increases, which decreases the probability of packet collisions. However, spectrum reuse is decreased as a result of the larger reserved area per transmission/reception.

\begin{it}Contention slots\end{it}: Each coordinator marks a few time slots as contention slots, in which nodes inside the cell (that are not currently scheduled for transmission) can send a request to initiate a new transmission. In the contention slots, nodes contend with each other using a CSMA MAC scheme to send a transmission request to their cell coordinators. Adjacent coordinators mark the same idle time slot(s) as contention slots. Coordinators dynamically adjust the number of contention slots and contention window size based on the traffic load condition. In Appendix, we present a mathematical model to calculate the number of successful transmission requests in the contention slots and the average delay to initiate a new transmission. Using the analytical model, we propose a mechanism to dynamically adjust the contention window size and the number of contention slots based to the network load and the required delay to initiate a new transmission.

\begin{figure*}
\centering
\subfigure[$r_a=r_g$]{\includegraphics[width=2.31in, trim=7cm 5cm 7cm 5cm, clip=true]{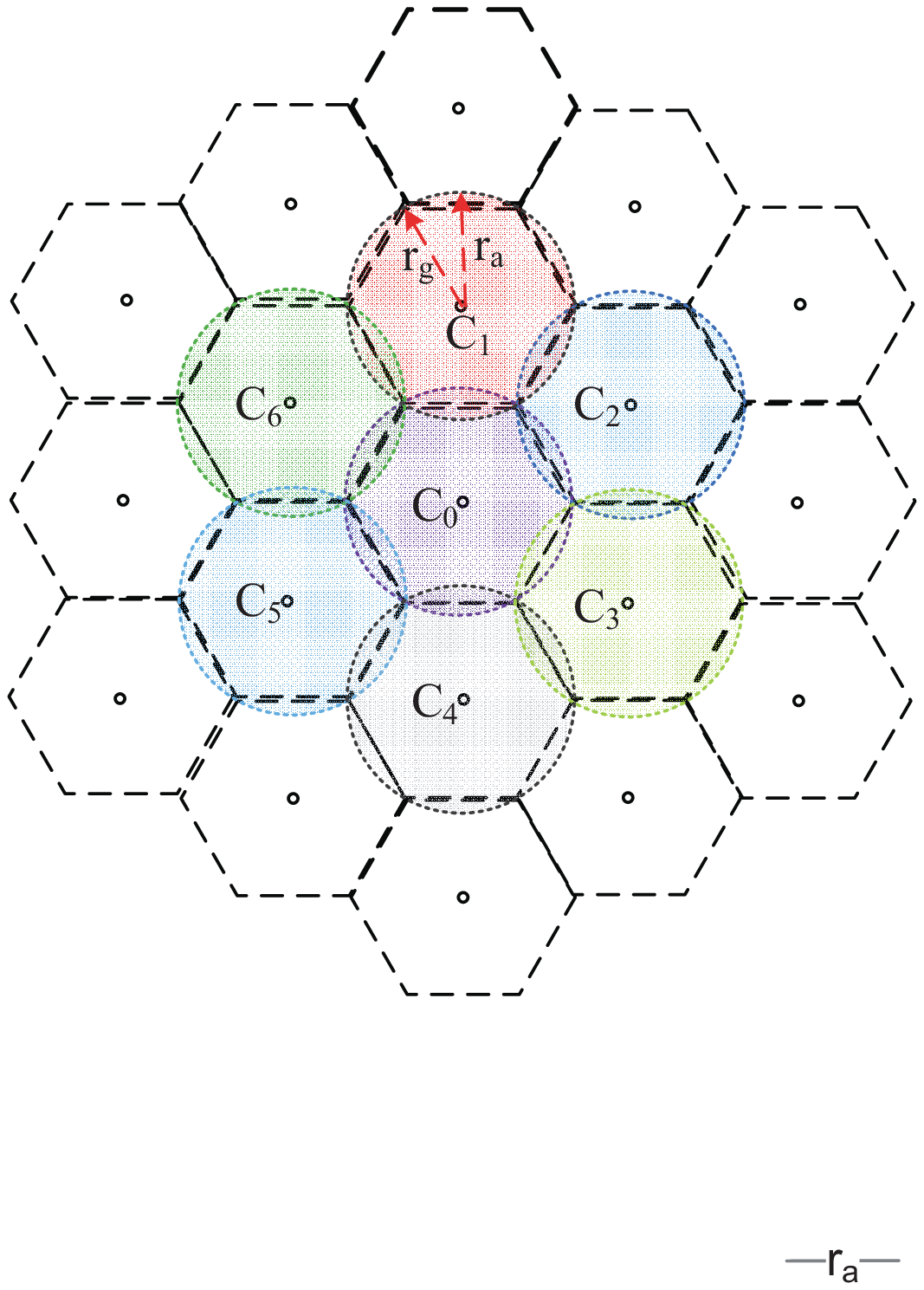}\label{InfR1}}
\subfigure[$r_a=1.5r_g$]{\includegraphics[width=2.31in, trim=7cm 5cm 7cm 5cm, clip=true]{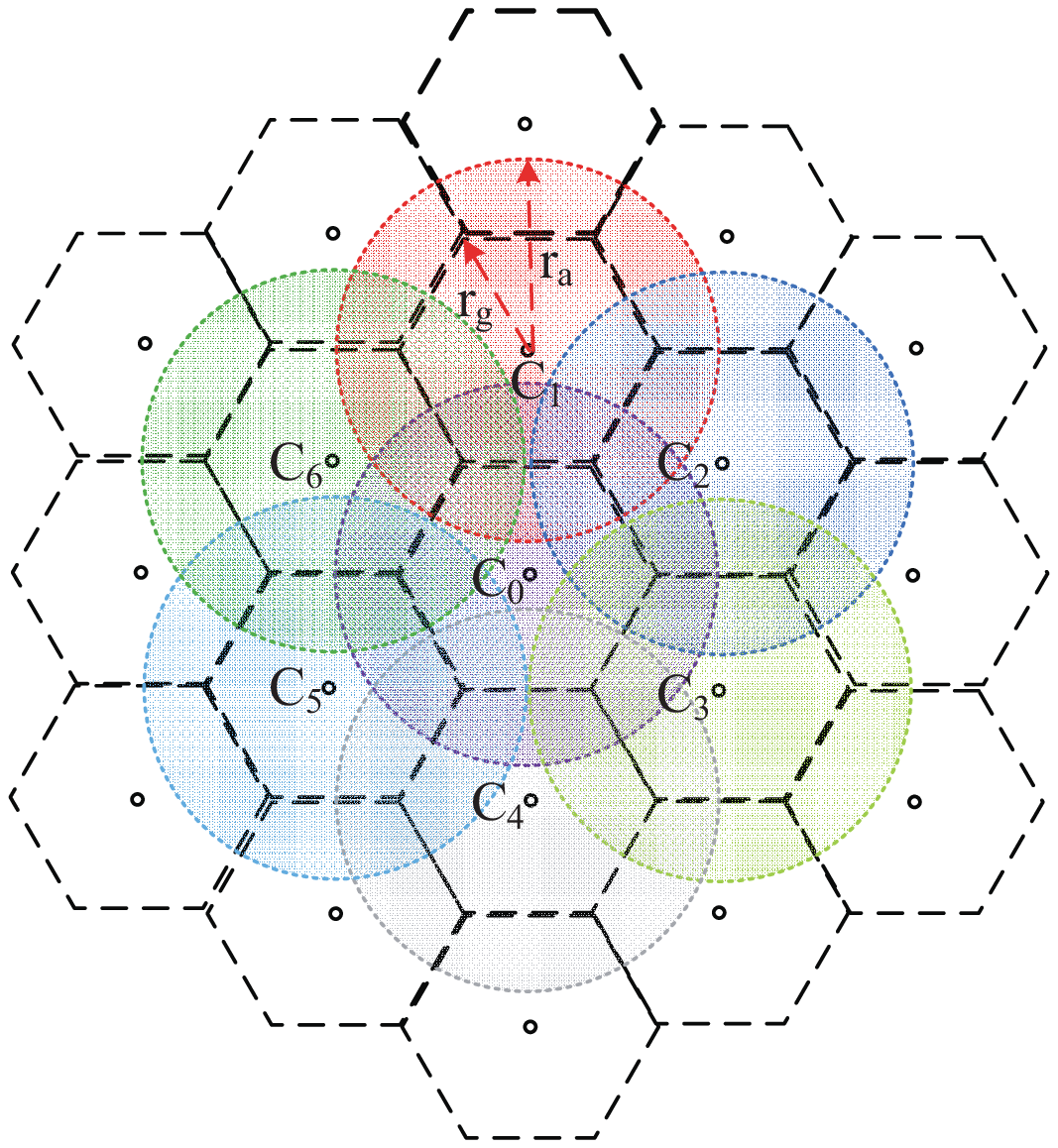}\label{InfR2} }
\subfigure[$r_a=2r_g$]{\includegraphics[width=2.31in, trim=7cm 5cm 7cm 5cm, clip=true]{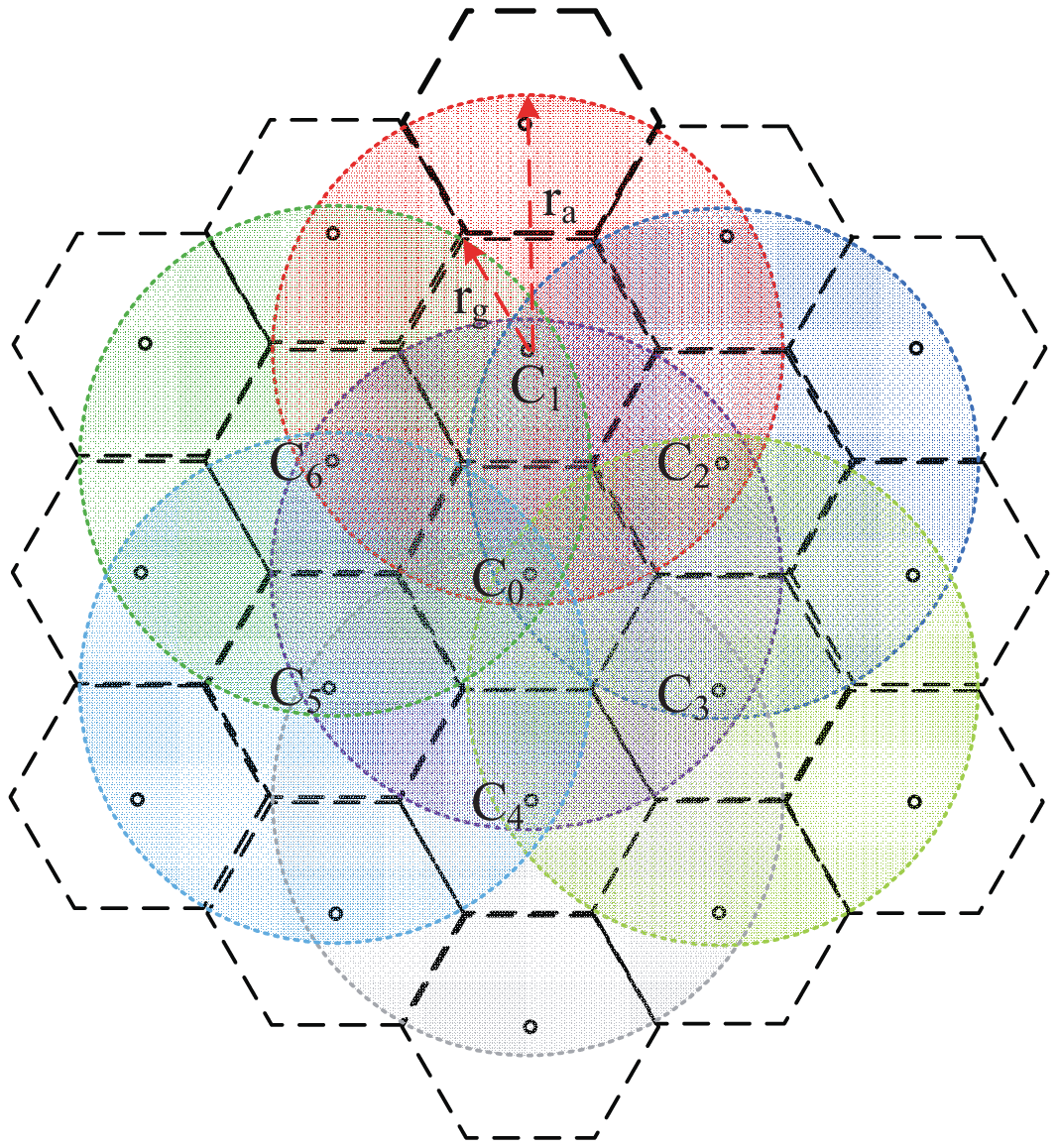} \label{InfR3}}
\caption{The area centred at coordinator $C_0$ in which the coordinator obtains the information of scheduled transmissions/receptions by overhearing scheduling packets of adjacent coordinators, where a circular area centred at each coordinator denotes the area that the coordinator broadcasts the information of scheduled transmissions/receptions.} \label{InfR}
\end{figure*}

\subsection{Operation of the MAC protocol}
A coordinator node stays awake during the following time slots in a frame:
\begin{enumerate}
  \item Scheduling slots -- to transmit a scheduling packet and to receive the scheduling packets transmitted by adjacent coordinators;
  \item One of the contention-free slot(s) scheduled for the transmission of each source inside the cell -- to receive the information of transmission request for the next frame, included in the header of a packet transmitted by the source node scheduled for transmission;
  \item Contention slots -- to receive transmission requests from nodes inside the cell that want to initiate a new transmission.
\end{enumerate}
Each coordinator has the location information of all nodes inside the cell and the nodes whose transmission/reception is advertised by adjacent coordinators. A coordinator maintains two tables:
\begin{enumerate}
  \item \begin{it}Demand table\end{it}, which contains the transmission requests of source nodes (i.e., source ID, destination ID, and the number of packets ready for transmission), and is updated/generated based on the nodes' transmission requests in previous frames and scheduling packets of adjacent coordinators;
  \item \begin{it}Scheduling table\end{it}, which contains the information of scheduled transmisssions/receptions (and correspondingly the reserved space for each scheduled transmission/reception) for the current frame, and is updated based on scheduling packets of coordinator and scheduling packets broadcasted by adjacent coordinators.
\end{enumerate}
Based on the demand table and scheduling table, each coordinator transmits a scheduling packet at its assigned scheduling time slot in each frame. The scheduling packet contains the following information:
\begin{enumerate}
  \item the schedule of transmissions/receptions (scheduled by the coordinator and/or adjacent coordinators) within distance $r_a$ of the coordinator, where $r_a\in[r_g, 2r_g]$;
  \item cancelation of scheduled transmissions/receptions by adjacent coordinators within distance $r_a$ of the coordinator that interfere with transmissions/receptions scheduled by other adjacent coordinators;
  \item announcement of the contention slots and contention window size for the current frame.
\end{enumerate}

Figure \ref{InfR} shows the area centred at a coordinator in which the coordinator obtains the information of scheduled transmissions/receptions by overhearing scheduling packets of adjacent coordinators. A coordinator will schedule a transmission from a source to a destination in a contention-free time slot only when neither an interfering node to the source is scheduled for reception nor an interfering node to the destination is scheduled for transmission. Also, each coordinator will cancel scheduled transmissions/receptions by adjacent coordinators within range $r_a$ that interfere with other existing scheduled transmissions/receptions. This mechanism ensures that a scheduled link for transmission/reception by a coordinator does not interfere with transmissions/receptions of nodes within range $r_a$ of the coordinator or adjacent coordinators. In Figure \ref{PMAC1}, a scheduled link for transmission by coordinator $C_0$ does not interfere with any other scheduled transmission/reception in area $A_0\cup A_1\cup....A_6$, where $A_i, i\in\{0,1,...6\}$ denote the area within range $r_a$ from coordinator $C_i$. To illustrate, consider frame $n$ where scheduling time slots are assigned as in Figure \ref{FrameStructureEx} and $C_{i}\in G_{i}$, $i\in\{0,1,2,...,6\}$. A transmission/reception scheduled by coordinator $C_0$ will not interfere with any scheduled transmission/reception in area $A_4$, $A_5$, $A_6$, and $A_0$, because coordinator $C_0$ receives the scheduling packets of $C_4$, $C_5$, and $C_6$ before transmitting its scheduling packet and it does not schedule an interfering transmission/reception. Also, coordinators $C_1$, $C_2$, and $C_3$, which overhear the scheduled transmission/reception from coordinator $C_0$ before transmitting their own scheduling packets, will not schedule an interfering transmission/reception and will cancel any interring transmission/reception scheduled by their adjacent coordinators in area $A_1$, $A_2$, and $A_3$ respectively.

When both source and destination nodes are in one cell, the cell coordinator finds time slot(s) to schedule contention-free transmission/reception and broadcast the scheduled transmission/reception in its scheduling time slot of current frame. However, when the source and destination nodes are located in adjacent cells, the coordinator of source schedules the transmission/reception in the current frame only if its scheduling time slot is before the scheduling time slot of the coordinator of destination node. Thus, the coordinator of destination node can inform the destination node of the scheduled transmission/reception in its scheduling time slot of the current frame. Otherwise, the coordinator of source node finds time slots to schedule contention-free transmission/reception in the next frame and includes the scheduled transmission/reception in its scheduling packet for the current frame. In the next frame, both the coordinators of source and destination again broadcast the scheduled transmission/reception in their scheduling time slots. Consider the network as illustrated in Figure \ref{PMAC1}, where scheduling time slots are assigned to coordinators as in Figure \ref{FrameStructureEx} and $C_{i}\in G_{i}$, $i\in\{0,1,2,...,6\}$. Coordinator $C_0$ can schedule transmission/reception between nodes $b$ and $c$ (that are inside the cell) in each frame and inform both source and destination in its scheduling time slot. Also, it can schedule transmission/reception from source node $c$ to destination node $d$ in frame $n$, in which coordinator $C_3$ can inform destination node $d$ of the scheduled transmission/reception in the same frame. However, coordinator $C_0$ will not schedule a transmission from source nodes $b$ to destination node $a$ in frame $n$, in which the scheduling time slot of coordinator $C_5$ comes before $C_0$. In frame $n$, coordinator $C_0$ finds time slots to schedule the transmission/reception (from source node $b$ to destination node $a$) for frame $n+1$ and includes the information in its scheduling packet of frame $n$. In frame $n+1$, both coordinators $C_0$ and $C_5$ broadcast the scheduled transmissions/receptions in their scheduling time slots.

\begin{figure}[t]
  \centering
  \subfigure[Coordinator]{\includegraphics[width=3.8in, trim=3cm 6.7cm 6.5cm 2.6cm, clip=true]{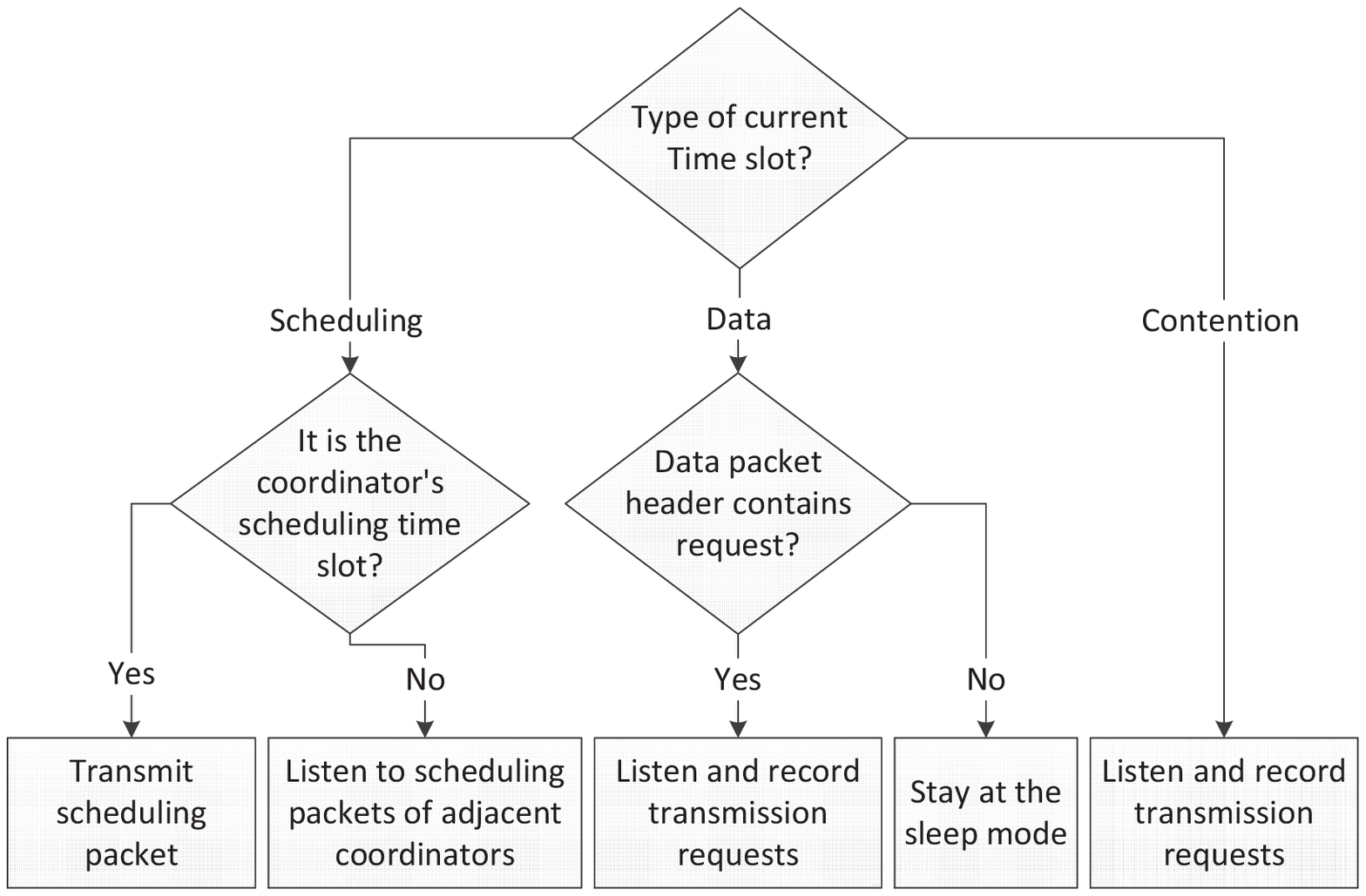}\label{FlowChart-Coordinator}}
  \\
  \subfigure[Non-coordinator]{\includegraphics[width=3.8in, trim=4cm 1.2cm 5.5cm 0.6cm, clip=true]{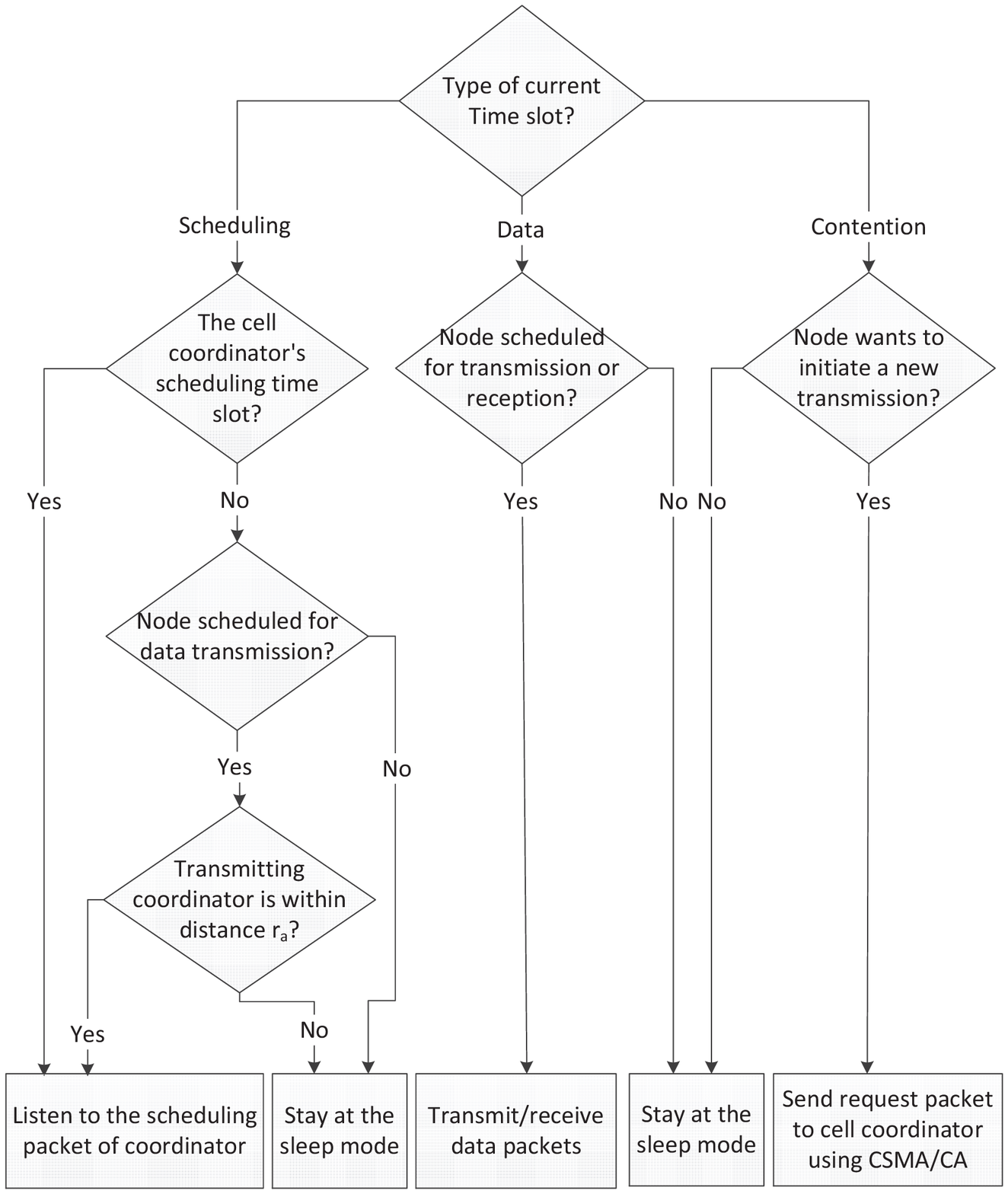}\label{FlowChart-Node}}
  \caption{The flowcharts of proposed MAC protocol in each time slot.}
  \label{FlowChart}
\end{figure}

Figure \ref{FlowChart} illustrates the operations of a coordinator node and a non-coordinator node in each time slot. Every non-coordinator node in the network stays awake during the scheduling time slot of its cell coordinator to receive the information of scheduled transmssions/receptions (in the contention-free slots) and contention slots in the current frame. A node scheduled for transmission will also stay awake during the scheduling time slots of the adjacent coordinators within distance $r_a$ from the node to receive cancelation information of transmission/reception (from adjacent coordinators). In Figure \ref{FrameStructureEx}, nodes $a$ and $b$ stay awake during scheduling time slot of coordinator $C_0$ in every time slot. Also, node $b$ stays awake during scheduling time slot of $C_5$ only if it is scheduled for transmission in the current frame. The source and destination nodes wake up at the assigned contention-free slots to perform transmissions/receptions as scheduled by cell coordinators. Source nodes will also include their transmission request for next frame in the header of one packet (as determined by cell coordinator). The cell coordinators will use this information to update its demand table for next frame. The source nodes that want to initiate a new transmission wake up at the assigned contention slots and contend with each other using a CSMA MAC scheme to send transmission request to cell coordinators. The coordinator will also record this information to update/generate its demand table for the next frame.

\section{Numerical Results}
Consider single-hop transmissions in a wireless ad hoc network with dimensions $6d_m\times6d_m$. Nodes are randomly distributed over the network coverage area and the destination of each source node is randomly selected from the rest nodes in its proximity at a distance less than $d_m$.

We compare performance of the proposed scheme with the IEEE 802.11 DCF scheme without power saving (hereafter referred to as DCF) and in power saving mode (hereafter referred to as PSM). Packets are generated according to a Poisson process at each source node. All control/scheduling packets (including RTS, ACK, ATIM, ATIM-Back, and scheduling packets) are transmitted at the control/scheduling channel rate ($R_s$) and all data packets are transmitted at the data channel rate ($R_d$). The required SINR at the destination for control/sceduling and data packets are $\Gamma_s=6$ dB and $\Gamma_d=9$ dB, $17$ dB respectively\footnote{The corresponding control/scheduling and data rates, according to data in \cite{Cisco2} for IEEE 802.11g, are $R_s=6$ Mbps and $R_d=18, 24$ Mbps respectively.}. The network load is defined as the aggregate packet generation rate in all the nodes. The following metrics are used as performance measures to compare the MAC schemes:
\begin{enumerate}
  \item \begin{it}Throughput\end{it}, which is defined as the summation of the numbers of packets transmitted per second from all nodes in network, weighted by the packet transmission distance;
  \item \begin{it}Energy consumption\end{it}, which is the average energy consumption per data packet, and is calculated as the ratio of total energy consumption in all nodes (including coordinators in our proposed scheme) to the total number of transmitted data packets in the network;
  \item \begin{it}Collision rate\end{it}, which is the ratio of collided data packets to the total number of transmitted data packets in the network.
\end{enumerate}
Similar metrics are used as performance measures in \cite{Kamal2, DPSM, TMMAC}, and \cite{On-demand, Kamal, NAPman, E-MiLi}. Each performance metric is calculated as the average performance over 10 different random node distributions in the network area. In our proposed MAC scheme, the network coverage area is partitioned into hexagon cells and a coordinator node is placed at the center of each cell as in Figure \ref{PMAC1}. We set $r_g=hd_m$ and $r_a=qr_g$, where $h\in\{1, 1.5, 2\}$ and $q\in\{1,1.2,...,2\}$. The frame duration is 100 ms and the duration of each scheduling, contention-free, and contention slot is 1 ms. The carrier sensing range during contention slots in the proposed scheme is set to $r_c=2r_g$. Since the performance of DCF in a wireless ad hoc network significantly depends on the carrier sensing range of the nodes, we vary carrier sensing range from $r_c=1.8 d_m$ to $r_c=3.0d_m$. The beacon interval size of PSM is set to 100 ms \cite{Standard}. The ATIM size varies from 2 ms to 10 ms, which include the 4ms as specified in \cite{Standard}. Simulations are performed using MATLAB for 20 seconds of the channel time. Other simulation parameters are given in Table \ref{Table2}.
\begin{table}[tb]
    \begin{center}
    \caption{Simulation Parameters}
     \label{Table2}
        \begin{tabular}{l|l}
          \hline
          \hline
          Parameter & Value  \\
           \hline
          Mini-slot &     $20$ $\mu$s  \\
          SIFS      &      $10$ $\mu$s   \\
          PHY preamble & $192$ $\mu$s   \\
          RTS size & $160$ bits \\
          CTS size & $112$ bits \\
          ACK size & $112$ bits \\
          ATIM size & $224$ bits \\
          ATIM-ACK size & $112$ bits \\
          $CW_{min}$ &    $15$    \\
          $CW_{max}$ &    $1023$  \\
          Scheduling size for one transmission &  $200$ bits \\
          Scheduling time slot & $1$ms\\
          Contention-free time slot & $1$ms\\
          Contention time slot & $1$ms\\
          Data packet+SIFS+ACK+DIFS duration & $1$ms\\
          \hline
          $P_d$ & $100$ mW\\
          $P_s$ & $100-180$ mW\\
          $c$ & $0.0001$\\
          $c'$ & $3$\\
          $\alpha$ & $3.4$\\
          $d_m$ & $20$\\
          $R_d$ & $18$ Mbps, $24$ Mbps\\
          $R_s$ & $6$ Mbps\\
          $\Gamma_d$ & $9$ dB, $17$ dB\\
          $\Gamma_s$ & $6$ dB\\
          \hline
          Beacon interval & $100$ ms\\
          Frame duration & $100$ ms \\
          Power consumption in sleep mode & $0.075$ W\\
          Power consumption in receive mode & $1.15$ W\\
          Power consumption in transmit mode& $2.25-3.15$ W\\
          \hline
           \hline
        \end{tabular}
    \end{center}
\end{table}

\begin{figure}
\centering
\subfigure[$\Gamma_d=9$ dB]{\includegraphics[width=2.71in, trim=.9cm 0cm 1.1cm 0cm, clip=true]{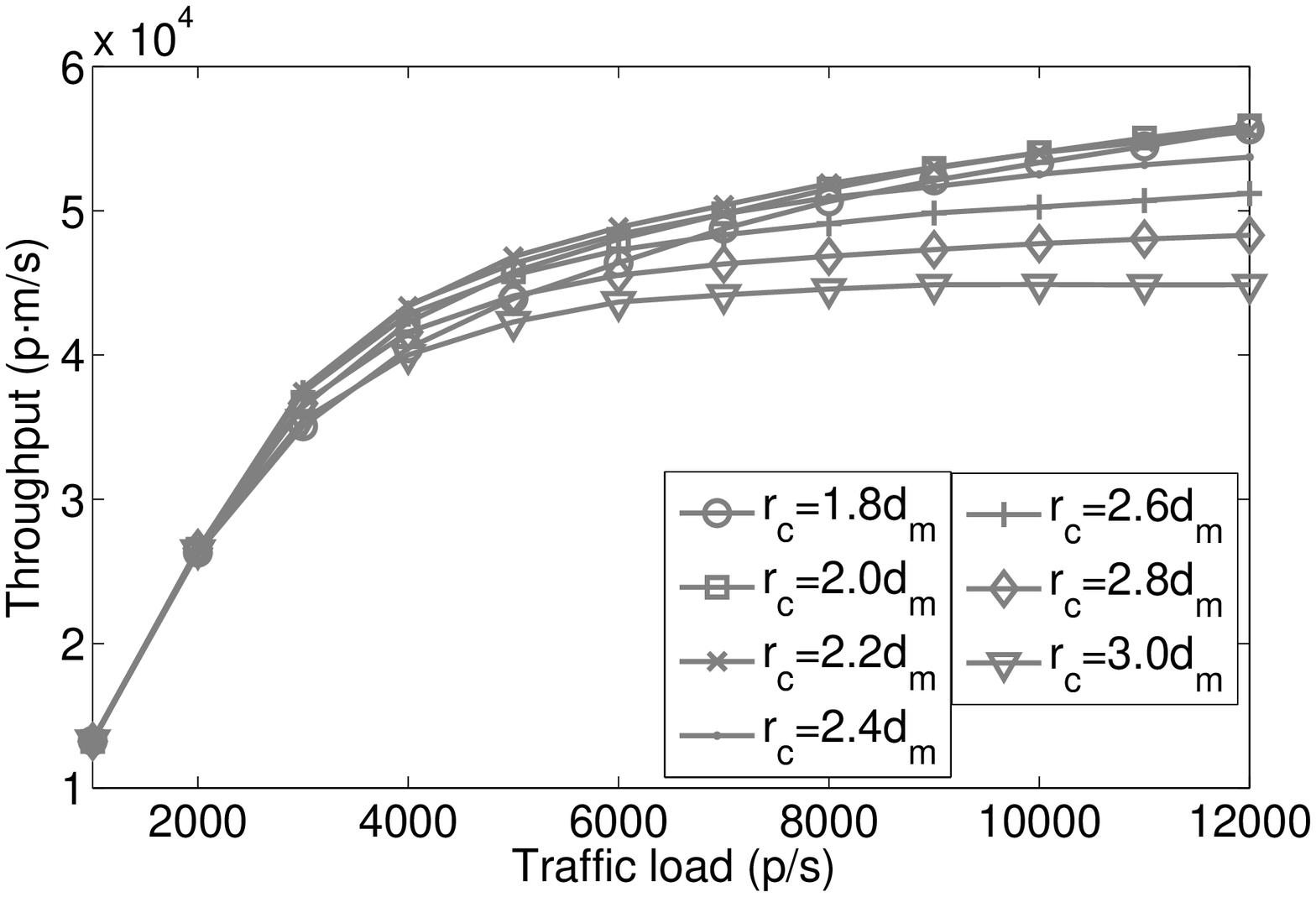}\label{DCF2} }
\subfigure[$\Gamma_d=17$ dB]{\includegraphics[width=2.71in, trim=.9cm 0cm 1.1cm 0cm, clip=true]{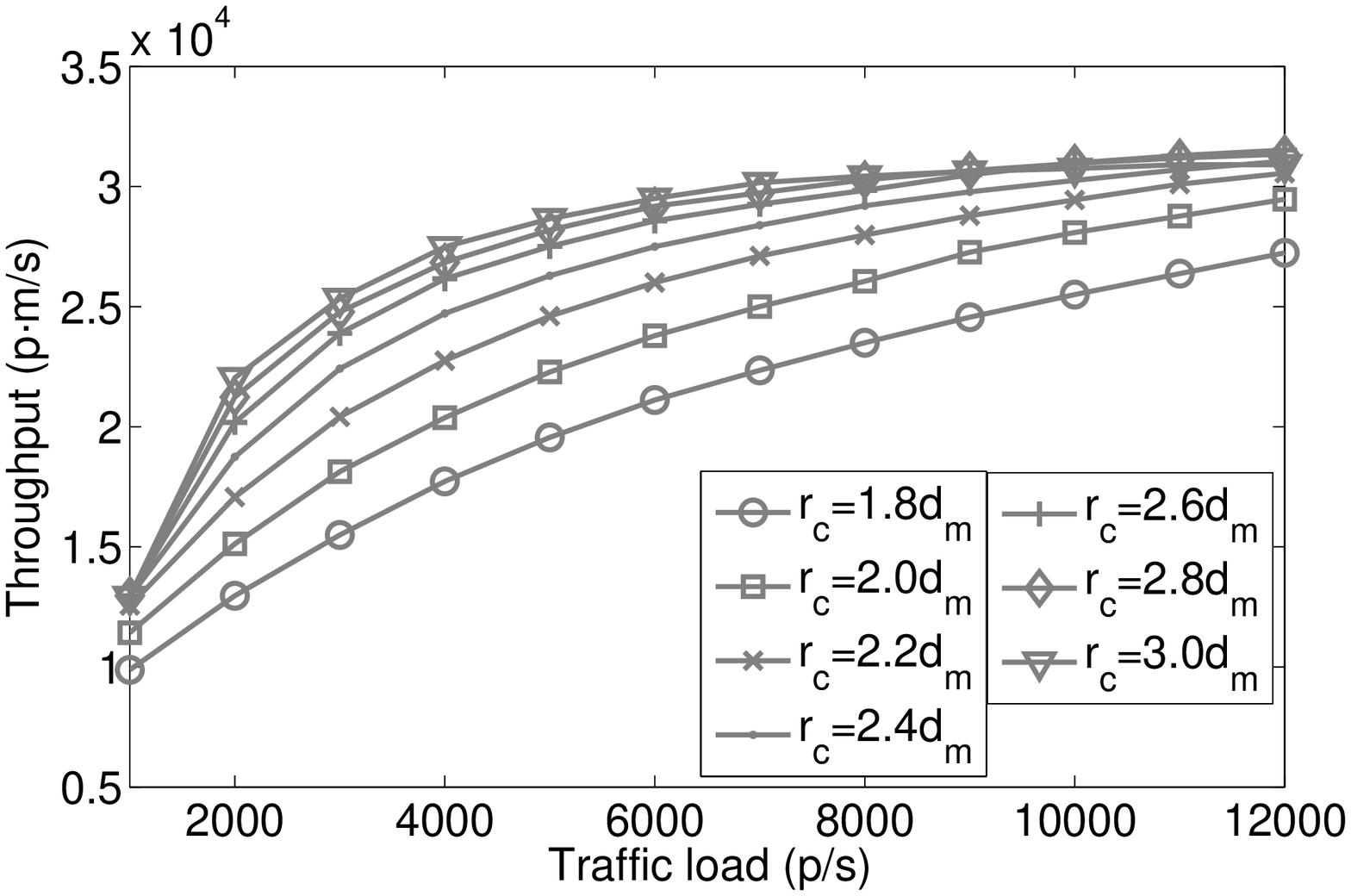} \label{DCF3}}
\caption{Throughput of the IEEE 802.11 DCF MAC vs traffic load for different carrier sensing ranges (N=100, $\Gamma_s=6$ dB).} \label{DCF}
\end{figure}

\begin{figure}
\centering
\subfigure[$\Gamma_d=9$ dB]{\includegraphics[width=2.71in, trim=.9cm 0cm 1.1cm 0cm, clip=true]{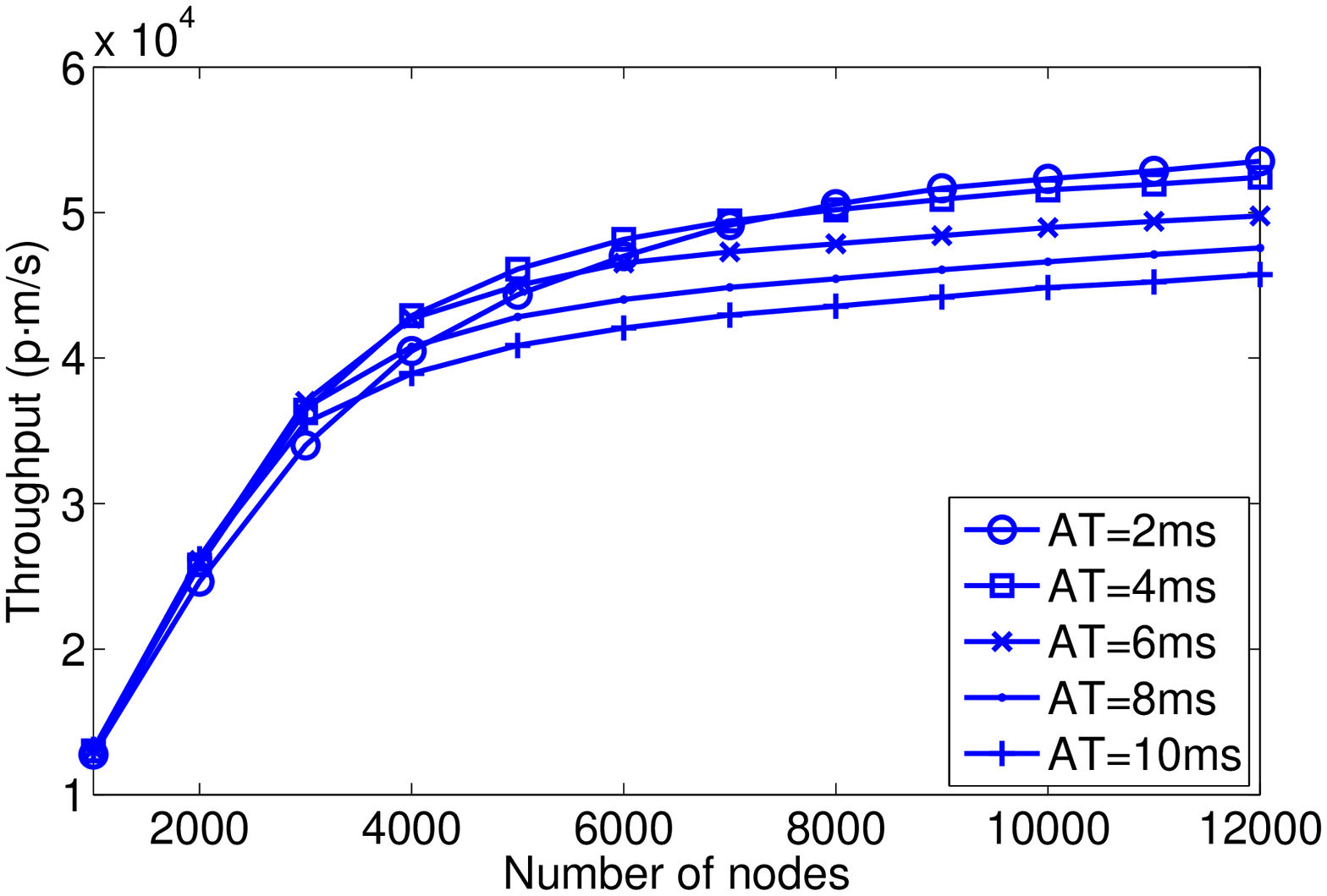}\label{PSM2} }
\subfigure[$\Gamma_d=17$ dB]{\includegraphics[width=2.71in, trim=.9cm 0cm 1.1cm 0cm, clip=true]{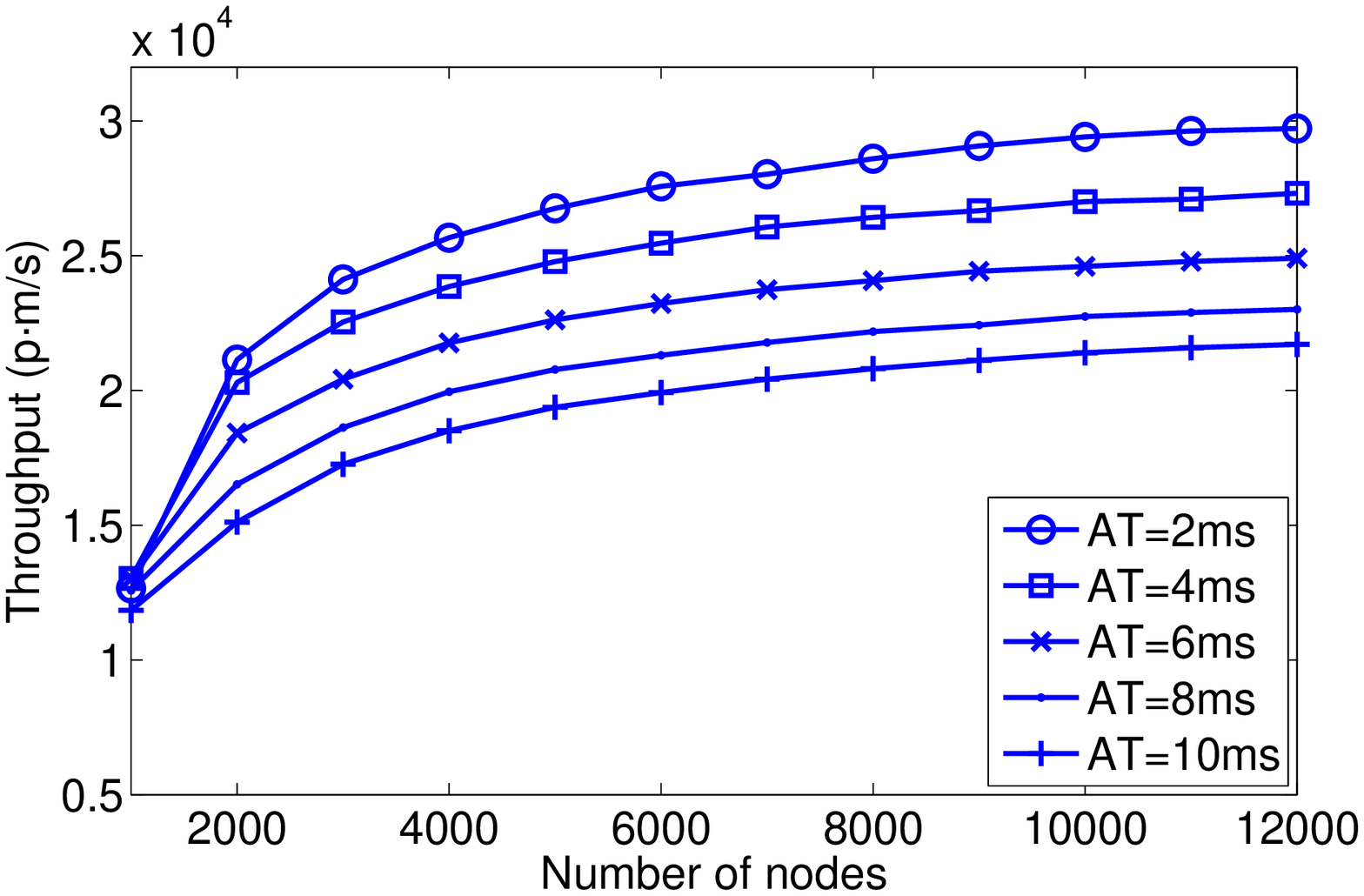} \label{PSM3}}
\caption{Throughput of the IEEE 802.11 DCF MAC in power saving mode (PSM) vs traffic load for different ATIM size when the carrier sensing range is set for highest throughput (N=100, $\Gamma_s=6$ dB).} \label{PSM}
\end{figure}

Figure \ref{DCF} shows the throughput of IEEE 802.11 DCF MAC scheme versus traffic load as the carrier sensing range changes from $1.8 d_m$ to $3.0d_m$. It is observed that the throughput of DCF can be maximized by choosing $r_c=2.0d_m$ and $r_c=2.8d_m$ when $\Gamma_d=9$ dB and $\Gamma_d=17$ dB respectively. Figure \ref{PSM} shows the performance of PSM as the ATIM size changes from 2 ms to 10 ms using carrier sensing range corresponding to the highest throughput of DCF in Figure \ref{DCF}. According to Figure \ref{PSM}, the optimal choice of ATIM size to maximize the throughput depends on the network traffic load and required SINR at the receiver, $\Gamma_d$. We consider a DCF scheme and a PSM scheme whose carrier sensing range and ATIM size are adjusted for highest throughput, referred to as best-DCF and best-PSM hereafter.


\begin{figure*}
\centering
\subfigure[]{\includegraphics[width=2.31in, trim=.5cm 0cm 1cm 0cm, clip=true]{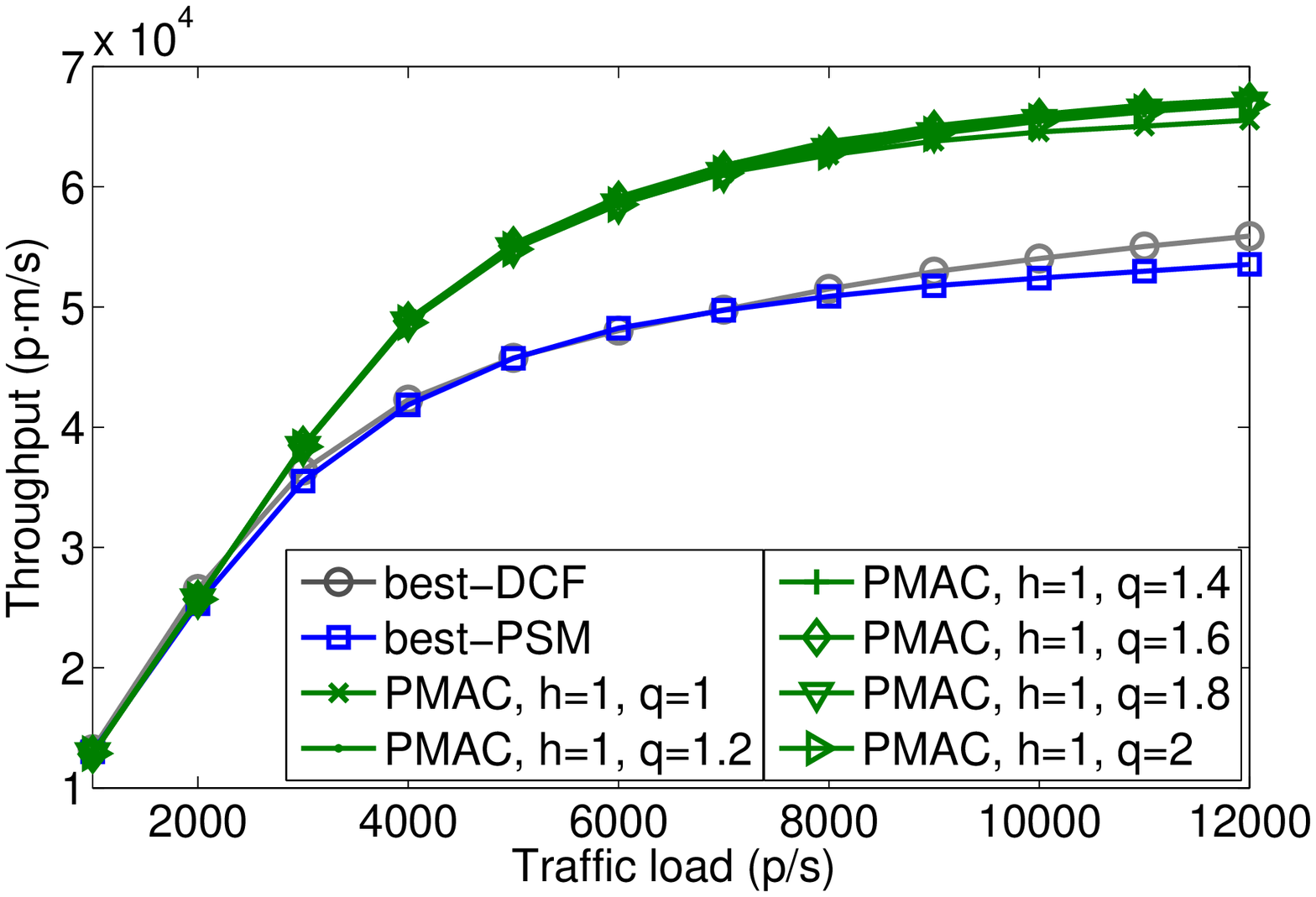}\label{Thpt9db-1}}
\subfigure[]{\includegraphics[width=2.31in, trim=.5cm 0cm 1cm 0cm, clip=true]{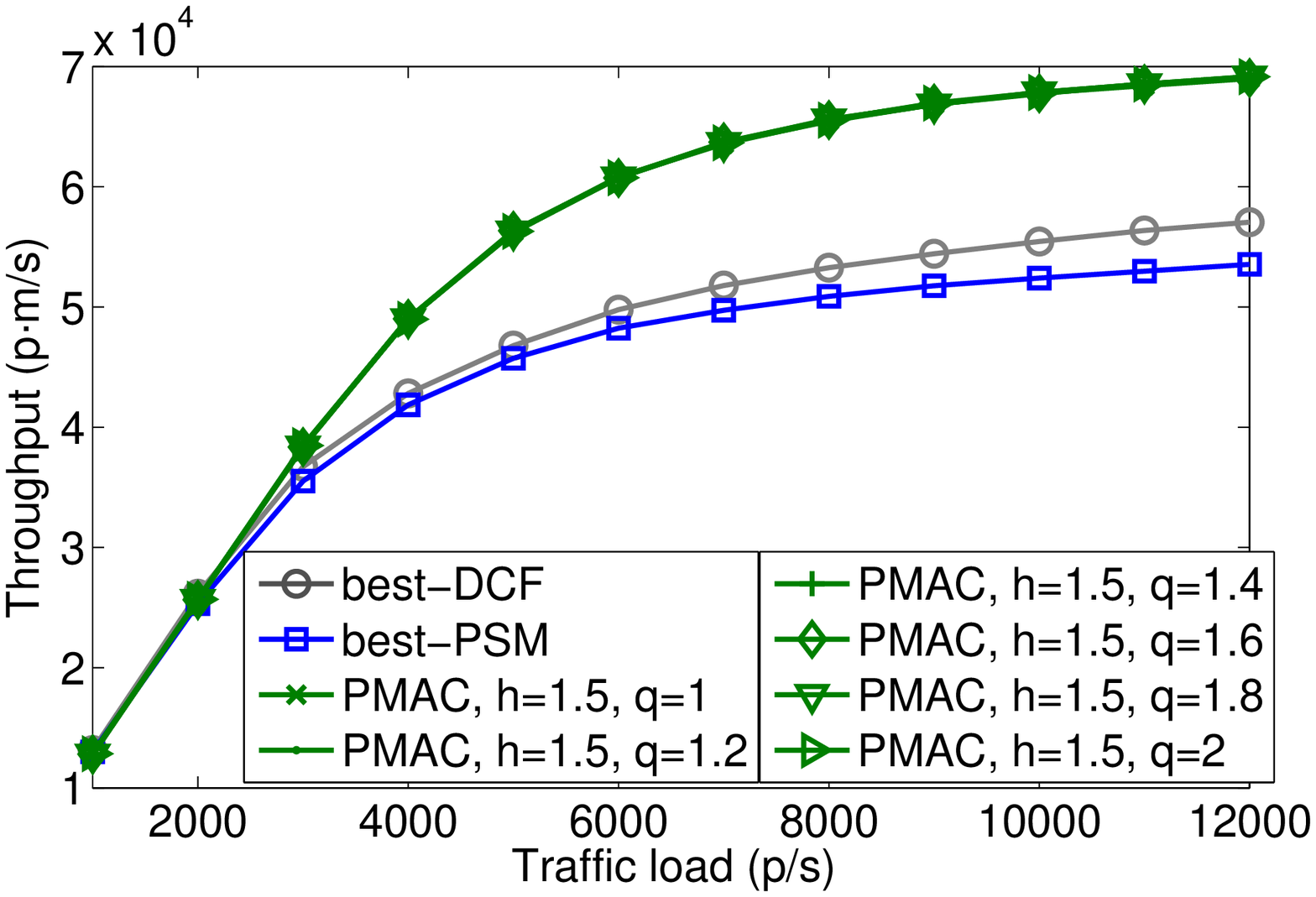}\label{Thpt9db-2} }
\subfigure[]{\includegraphics[width=2.31in, trim=.5cm 0cm 1cm 0cm, clip=true]{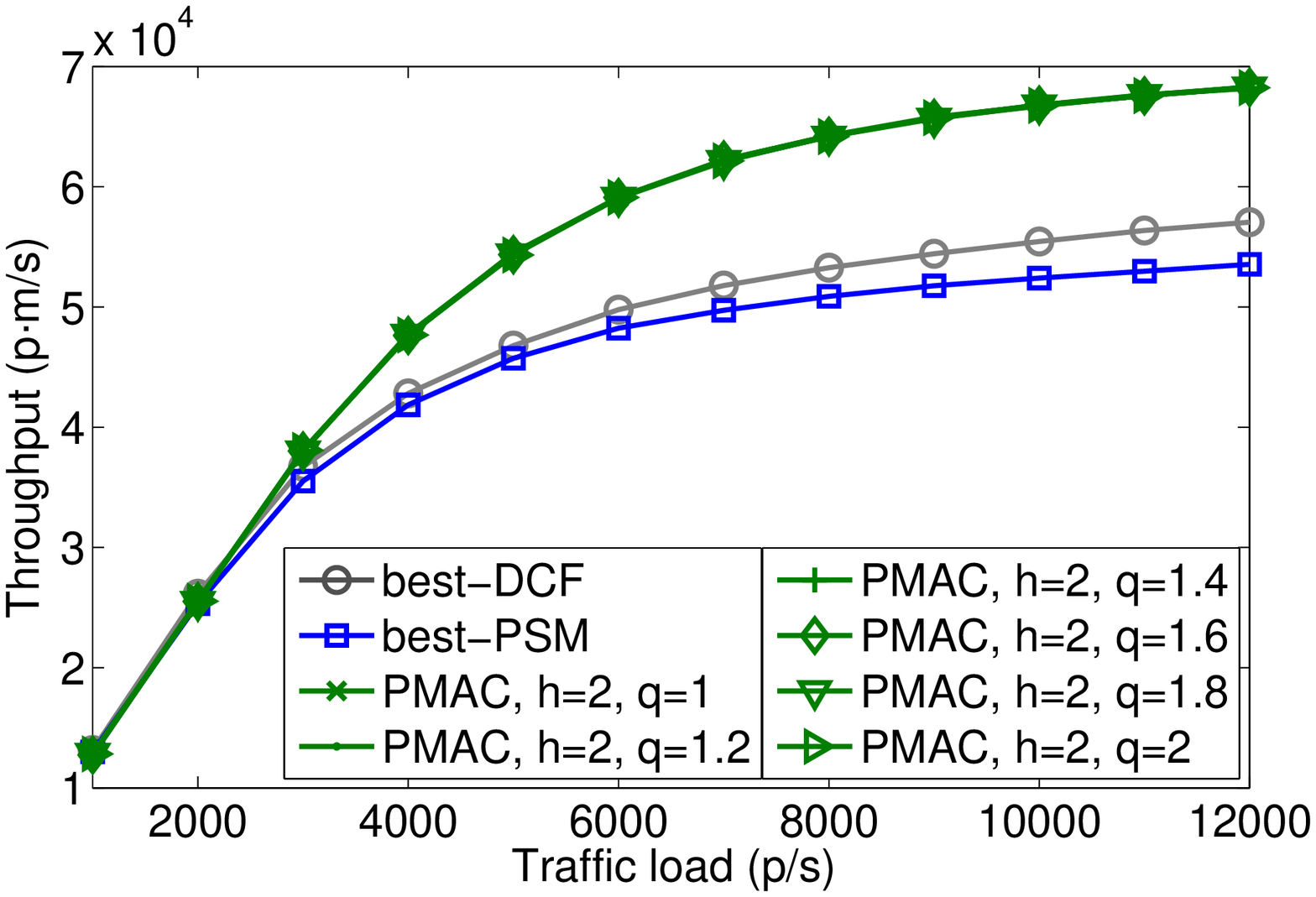} \label{Thpt9db-3}}
\caption{Throughput of the proposed MAC (PMAC), best-DCF, and best-PSM (N=100, $\Gamma_s=6$ dB and $\Gamma_d=9$ dB).} \label{Thpt9db}
\end{figure*}

\begin{figure*}
\centering
\subfigure[]{\includegraphics[width=2.31in, trim=.5cm 0cm 1cm 0cm, clip=true]{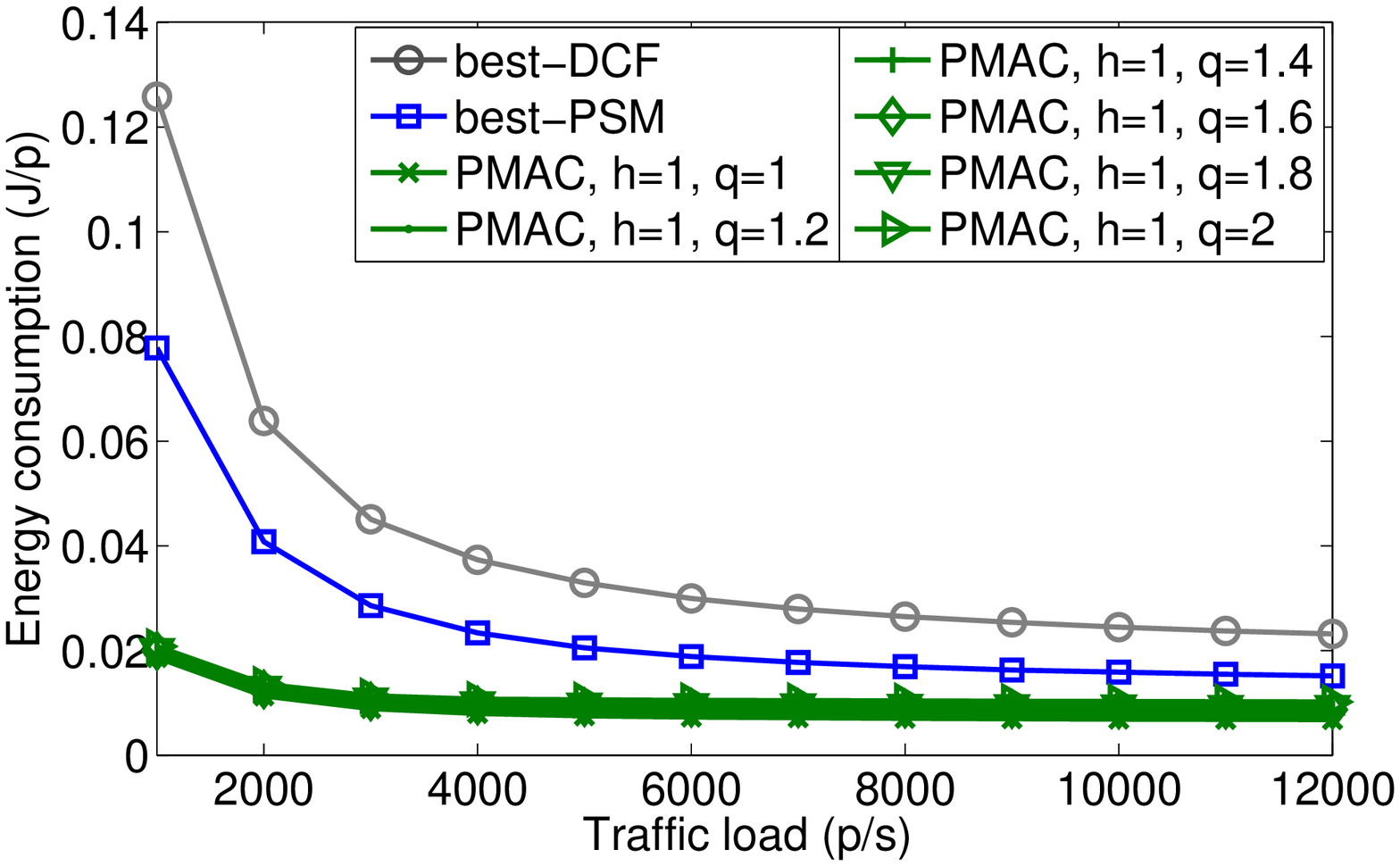}\label{ECPO9db-1}}
\subfigure[]{\includegraphics[width=2.31in, trim=.5cm 0cm 1cm 0cm, clip=true]{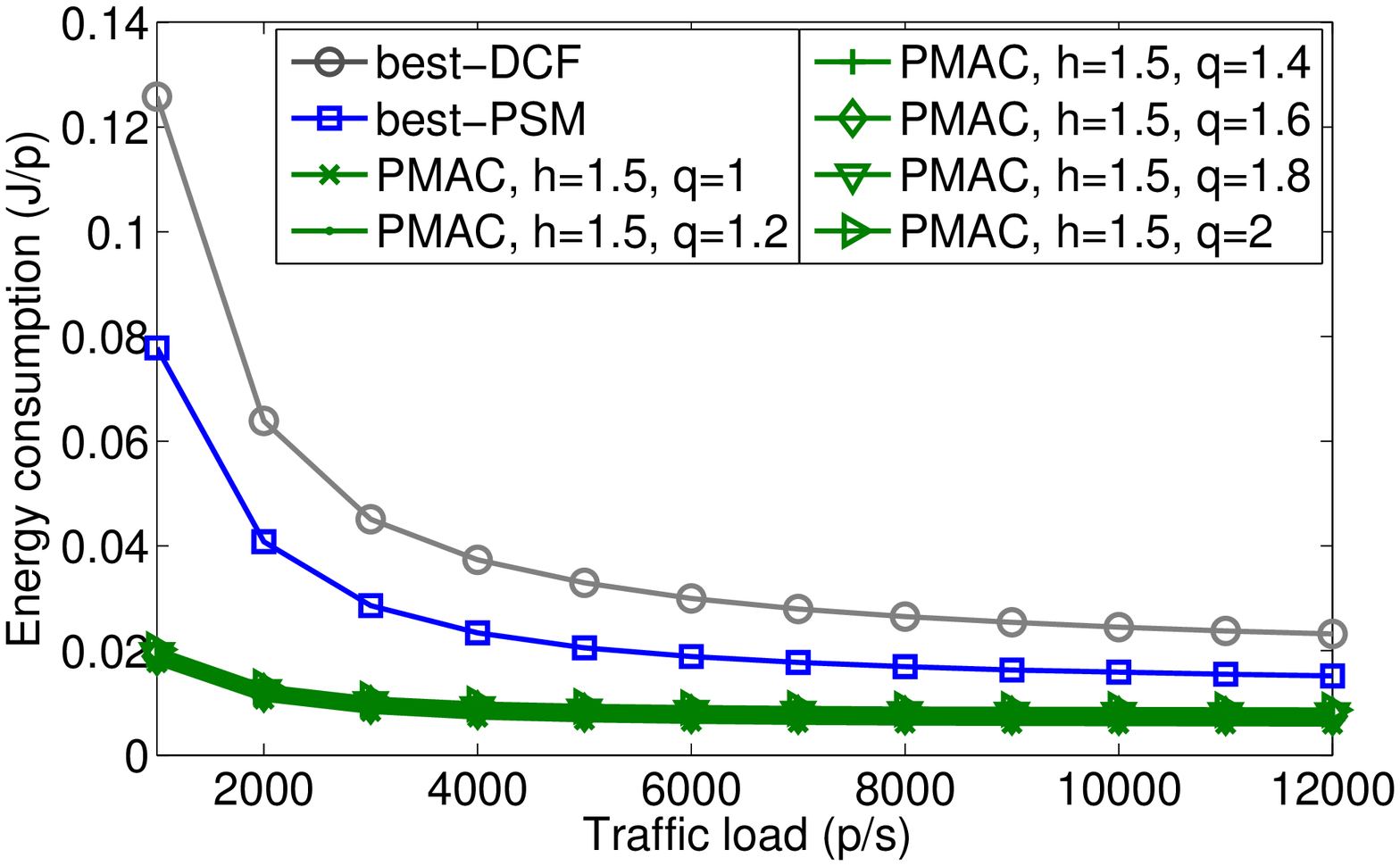}\label{ECPO9db-2} }
\subfigure[]{\includegraphics[width=2.31in, trim=.5cm 0cm 1cm 0cm, clip=true]{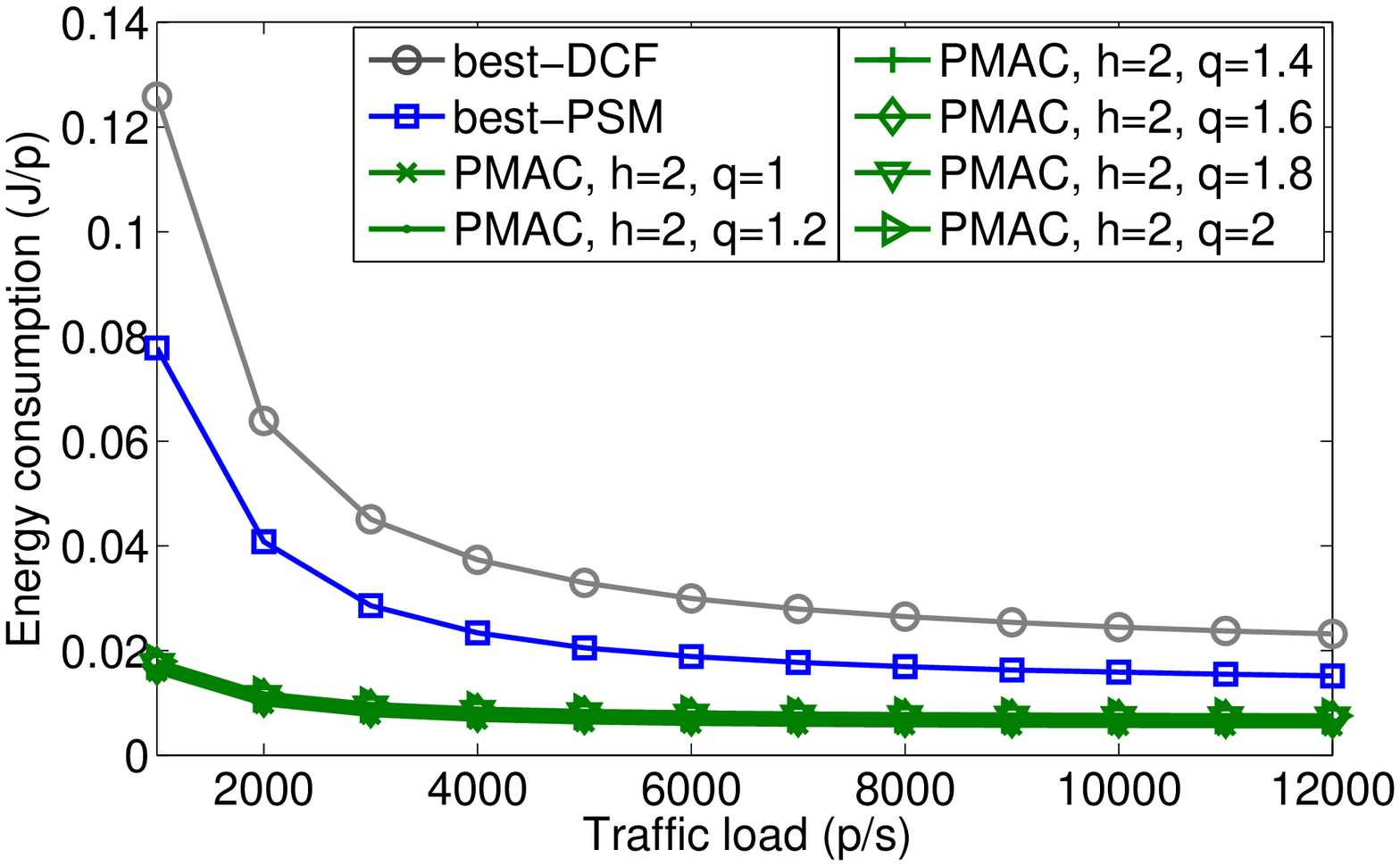} \label{ECPO9db-3}}
\caption{Energy consumption of the proposed MAC (PMAC), best-DCF, and best-PSM (N=100, $\Gamma_s=6$ dB and $\Gamma_d=9$ dB).} \label{ECPO9db}
\end{figure*}

\begin{figure*}
\centering
\subfigure[]{\includegraphics[width=2.31in, trim=.5cm 0cm 1cm 0cm, clip=true]{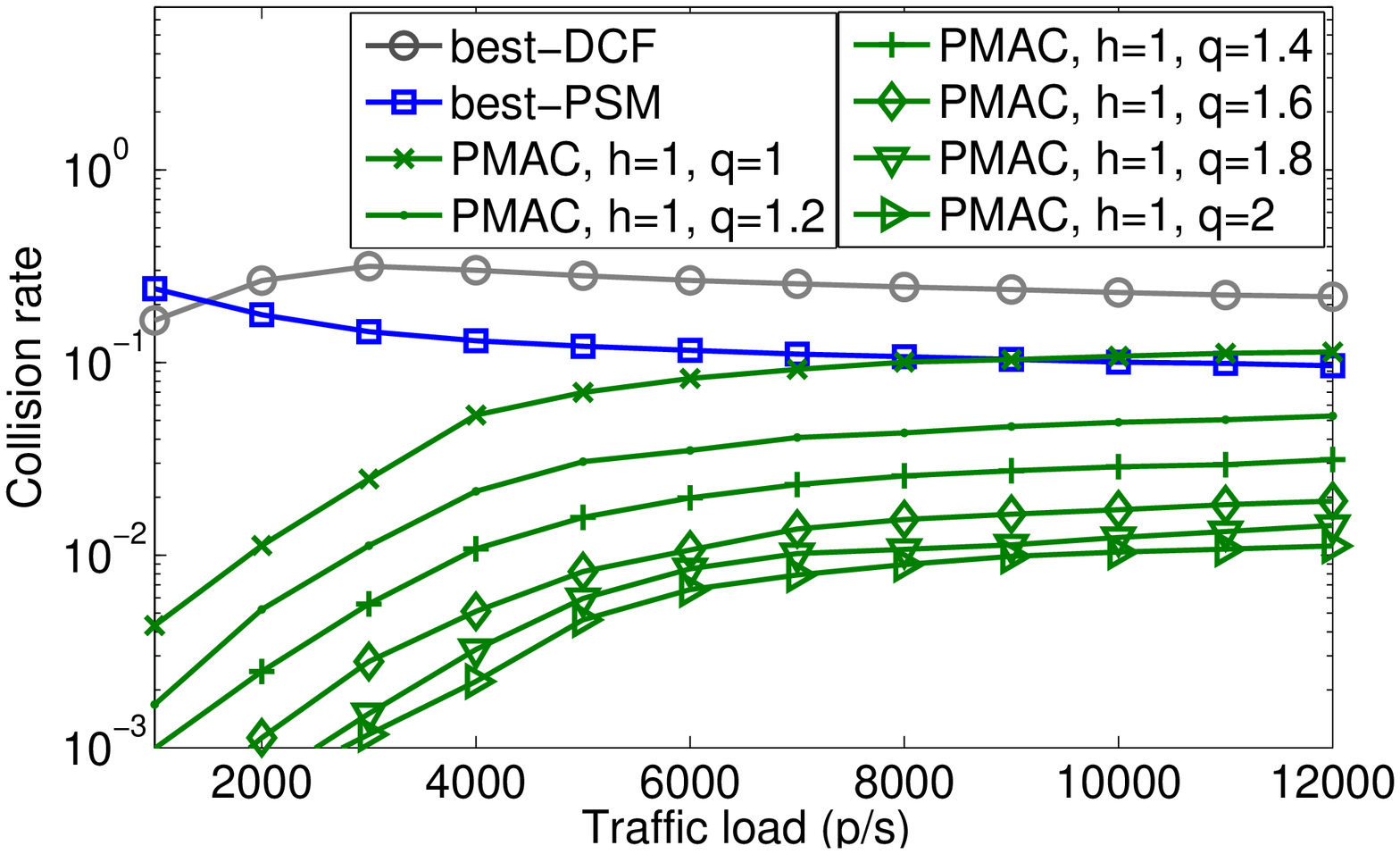}\label{ColiR9db-1}}
\subfigure[]{\includegraphics[width=2.31in, trim=.5cm 0cm 1cm 0cm, clip=true]{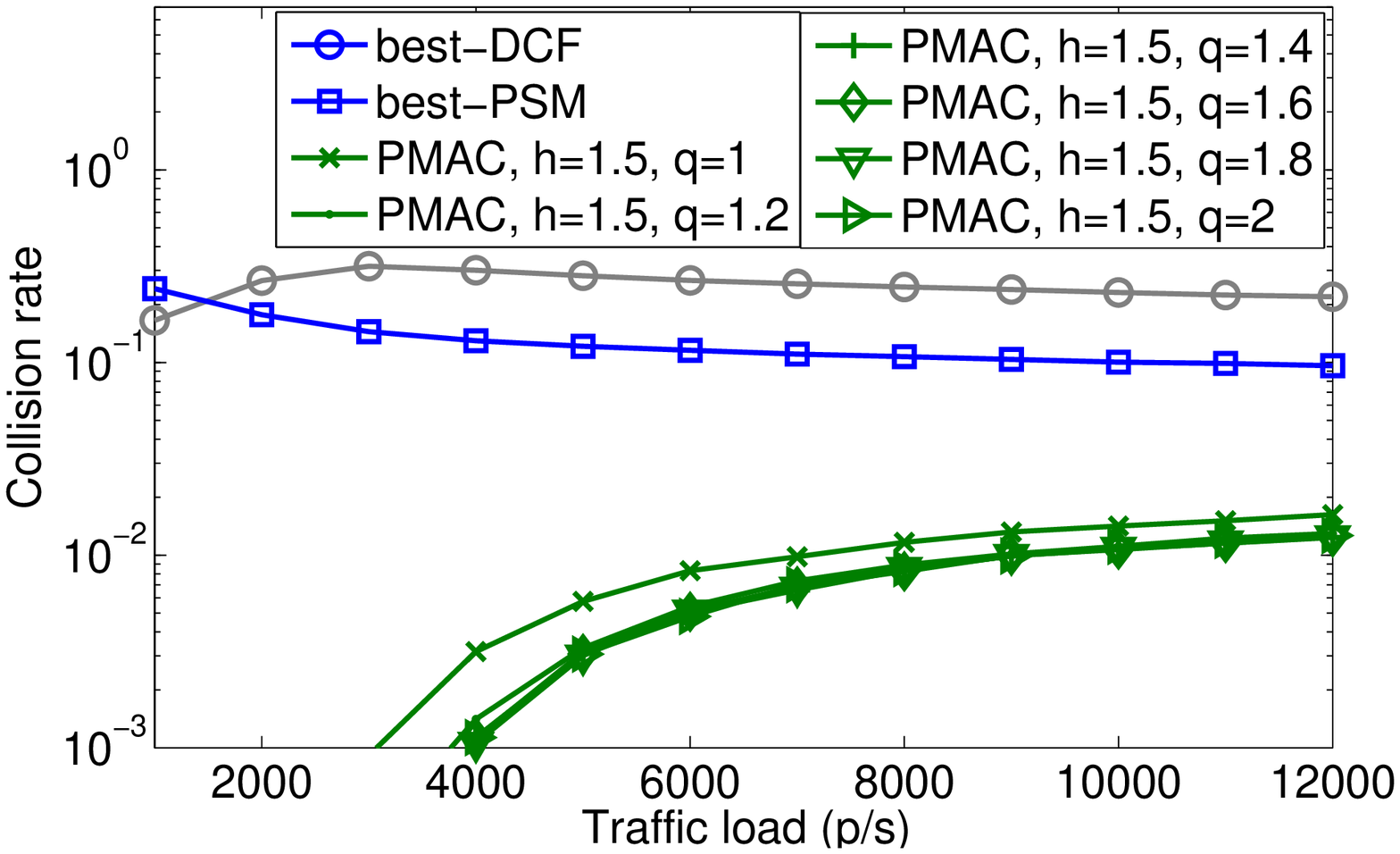}\label{ColiR9db-2} }
\subfigure[]{\includegraphics[width=2.31in, trim=.5cm 0cm 1cm 0cm, clip=true]{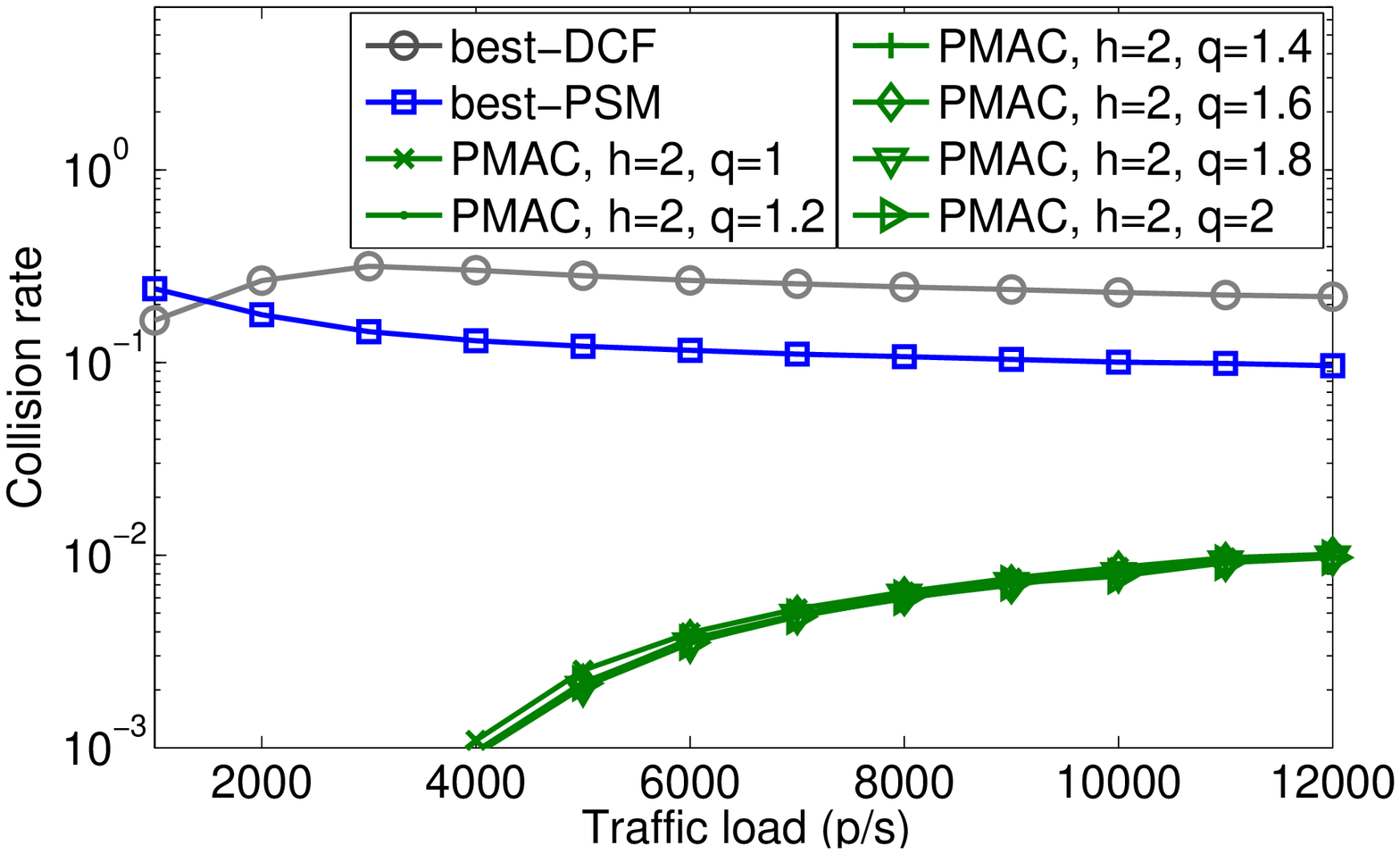} \label{ColiR9db-3}}
\caption{Collision rate of the proposed MAC (PMAC), best-DCF, and best-PSM (N=100, $\Gamma_s=6$ dB and $\Gamma_d=9$ dB).} \label{ColiR9db}
\end{figure*}

Figures \ref{Thpt9db}-\ref{ColiR9db} show the throughput, energy consumption and collision rate of the proposed MAC (PMAC), best-DCF, and best-PSM versus traffic load when $\Gamma_s=6$ dB and $\Gamma_d=9$ dB. From Figure \ref{Thpt9db}, the proposed MAC provides 20\% higher throughput than best-DCF and best-PSM. The proposed MAC mechanism can achieve high throughput by opportunistically utilizing the spectrum in space and time domains and reducing signaling overhead. Reserving the required space for each transmission/reception and sharing the information of scheduled transmissions among adjacent coordinators facilitate efficient spatial channel reuse, while avoiding transmission collisions, which significantly improve the network throughput. In addition, a cell coordinator schedules all data transmissions/receptions for nodes inside the cell by transmitting only a scheduling packet in each frame. The small scheduling overhead allows more data packet transmissions/receptions to increase throughput.


Energy consumption per transmitted data packet is shown in Figure \ref{ECPO9db}. Although the total energy consumption in each scheme increases as the network load increases, the highest energy consumption per packet occurs at the lowest network traffic load. The results indicate that the proposed MAC has significantly lower energy consumption per transmitted data packet, which is 25\%-50\% of the best-PSM energy consumption. The high energy efficiency of the proposed MAC scheme is the result of minimizing energy wastage because of node idle listening and transmission collisions, which is achieved by periodic assignment of deterministic time slots for transmissions/receptions. In the proposed scheme, a node stays awake only during the scheduling time slot of cell coordinator, in its data time slot(s) either for transmission or reception, and when initiating a new transmission in the contention slots. Also, energy wastage caused by transmission collisions is minimized by reserving space exclusively for each scheduled data transmission/reception and sharing the scheduling information among adjacent cell coordinators.

The packet collision rate for the different protocols is demonstrated in Figure \ref{ColiR9db}. The high transmission collision rate in the DCF and PSM MAC schemes is due to the hidden terminal problem of CSMA MAC in a wireless ad hoc network. In the proposed MAC, the packet collision rate is reduced as $r_a$ and/or $r_g$ increases, which increases the area range around a coordinator that it is aware of scheduled transmissions/receptions. As the results indicate, the proposed MAC has a much lower packet collision rate in compassion with the best-DCF and best-PSM. The proposed MAC scheme can effectively minimize transmission collisions by assigning contention-free time slots for data transmissions/receptions and reserving space around a scheduled link to prevent collisions.

\begin{figure*}
\centering
\subfigure[Throughput]{\includegraphics[width=2.31in, trim=.5cm 0cm 1cm 0cm, clip=true]{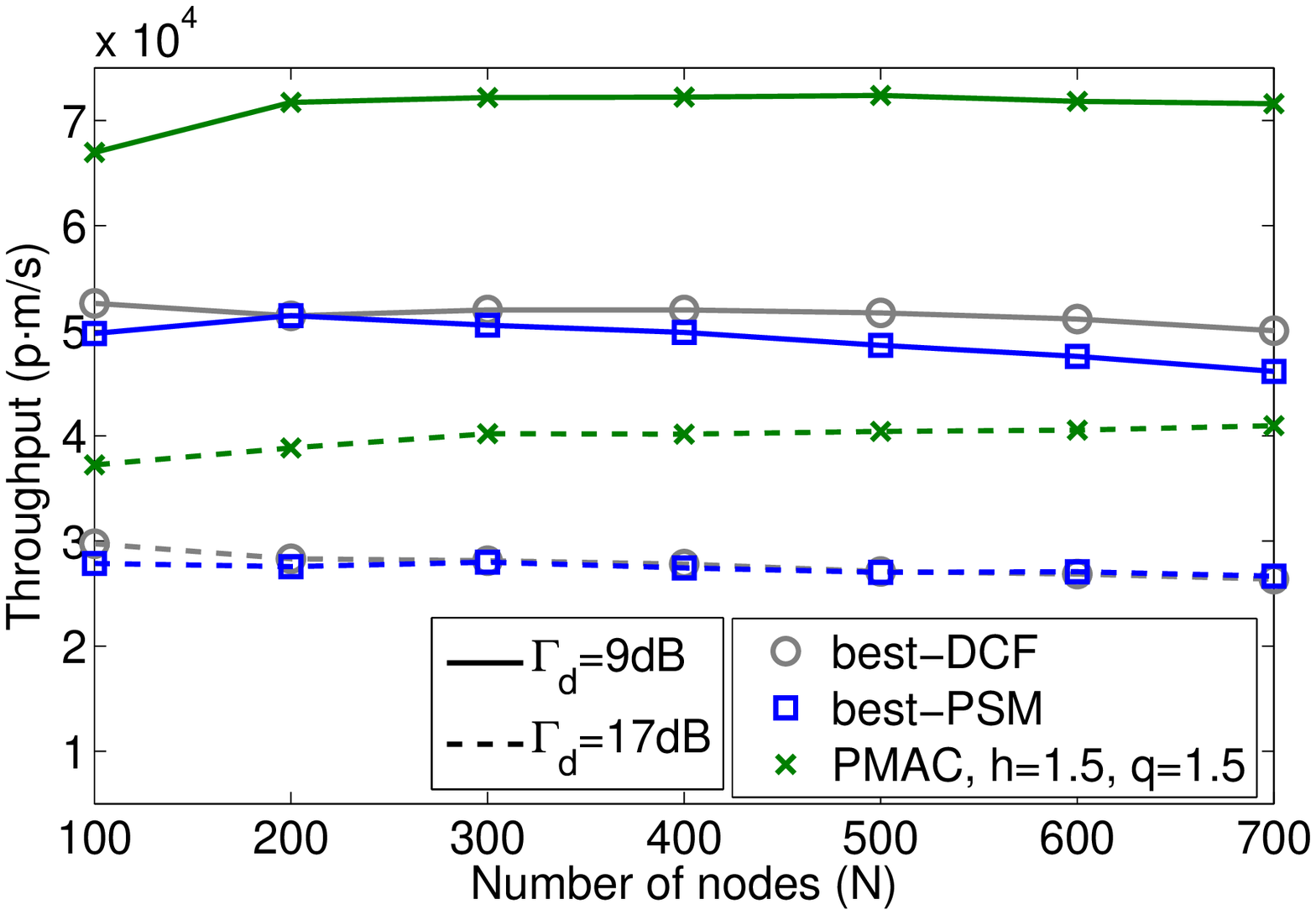}\label{Per_N-1}}
\subfigure[Energy consumption]{\includegraphics[width=2.31in, trim=.5cm 0cm 1cm 0cm, clip=true]{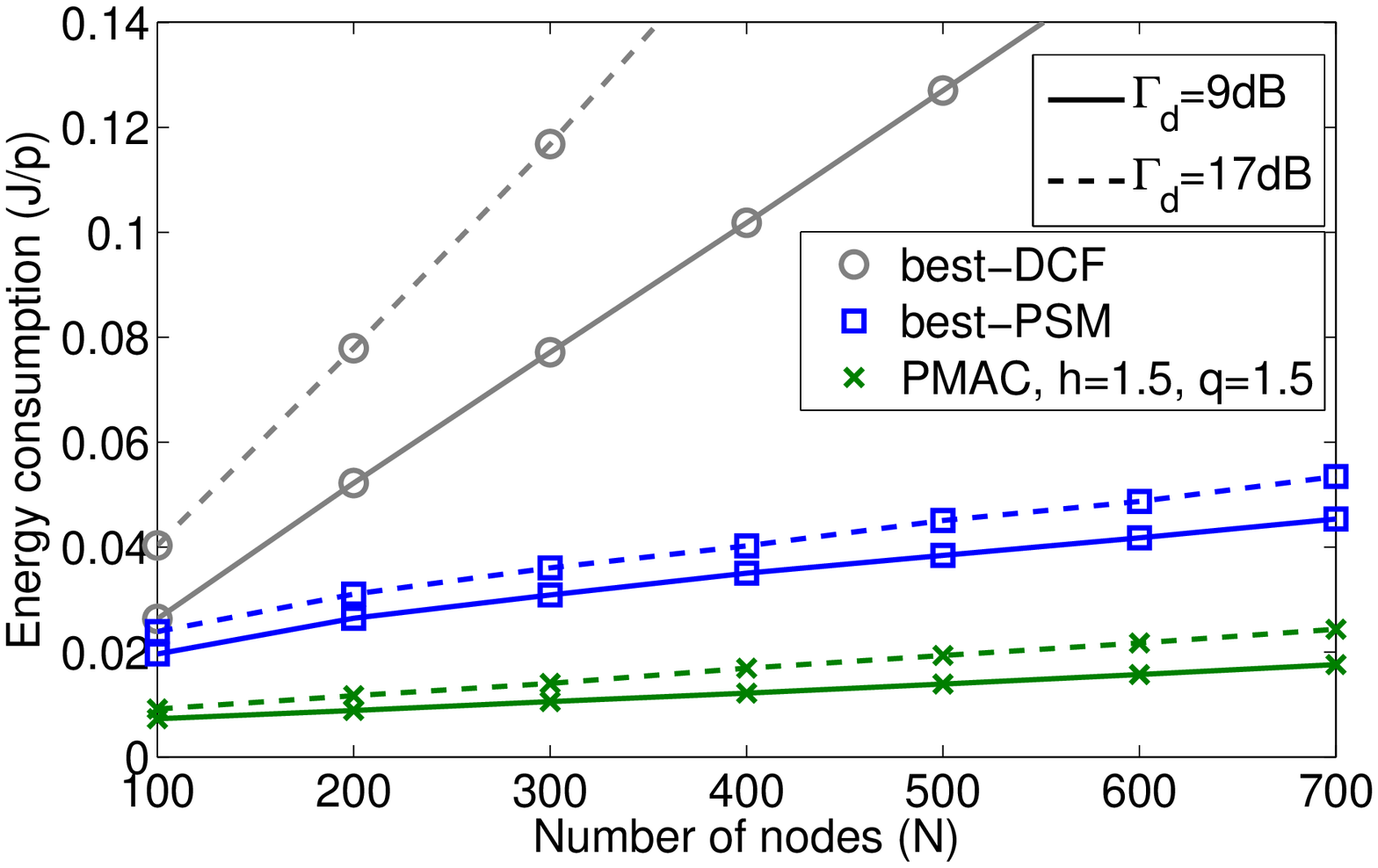}\label{Per_N-2} }
\subfigure[Collision rate]{\includegraphics[width=2.31in, trim=.5cm 0cm 1cm 0cm, clip=true]{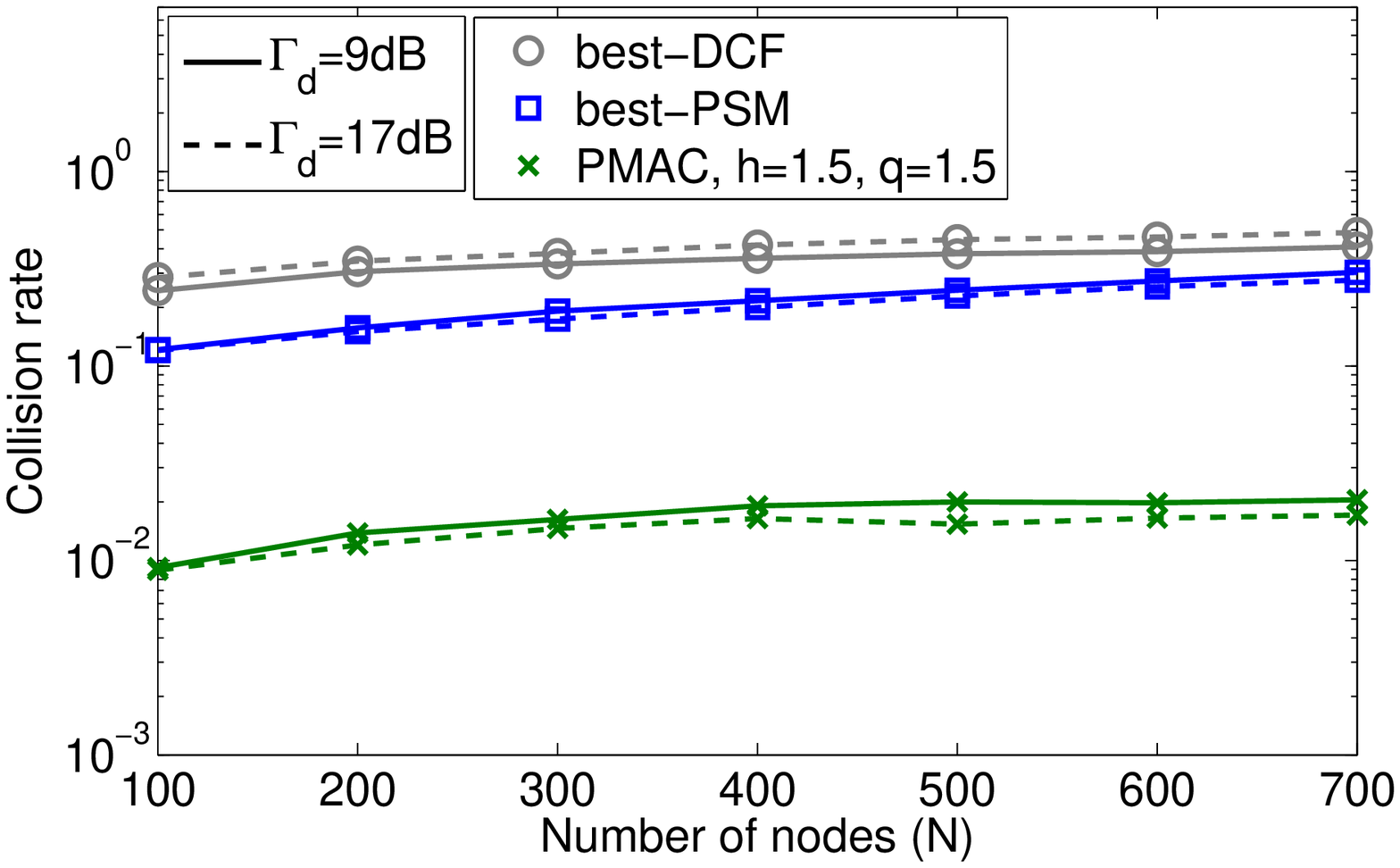} \label{Per_N-3}}
\caption{Performance of the proposed MAC (PMAC), best-DCF, and best-PSM versus node density (Traffic load=8000 p/s, $\Gamma_s=6$ dB and $\Gamma_d=9$, $17$ dB).} \label{Per_N}
\end{figure*}

Figure \ref{Per_N} shows the performance of the proposed MAC (PMAC), best-DCF, and best-PSM in high traffic load (8000 packets/s) as the node density changes and for $\Gamma_d=9$ dB and $17$ dB respectively. The number of transmitted packets per second decreases in each scheme as $\Gamma_d$ increases, because a larger channel space required for each transmission/reception in each MAC scheme to meet the higher SINR requirement at the receiver node. According to Figure \ref{Per_N-1}, the proposed MAC scheme provides 30\%-40\% higher throughput than best-DCF and best-PSM. Figure \ref{Per_N-2} shows that the energy consumption per packet increases in each scheme as the node density and/or $\Gamma_d$ increases. It is observed that the energy consumption of the proposed MAC mechanism is about 35\%-45\% of the best-PSM. Figure \ref{Per_N-3} shows that the transmission collision rate in the proposed MAC scheme is always lower than 0.02, which is about 10 times smaller than the transmission collision rate in the best-DCF and best-PSM.

\section{Conclusion}
In this paper, we present a novel coordination-based MAC protocol for a wireless ad hoc network. In the proposed MAC scheme, the network area is partitioned into cells and a coordinator node periodically schedules all transmissions/receptions for nodes inside its cell. For each scheduled transmission/reception, the channel in both time and space domains are reserved to avoid transmission collisions. Adjacent coordinators exchange scheduling information to maximize spatial spectrum reuse while avoiding transmission collision. A source node contends only once to transmit a batch of packets. After that it can request for transmission by including the information in the header of one data packet. Moreover, periodic scheduling of transmission time slots for data packets allows a node to put its radio interface into the sleep mode when not transmitting/receiving a packet in order to reduce energy consumption. We compare the performance of the proposed scheme with the IEEE 802.11 DCF scheme without power saving and in power saving mode, whose carrier sensing range and ATIM size are dynamically adjusted to provide highest throughput. The performance measures include aggregate throughput, average energy consumption per packet and packet collision rate. Simulation results show that the proposed scheme achievers substantially higher throughput, significantly reduces energy consumption, and has a much smaller packet collision rate in comparison with the existing protocols. Distributing coordinators in the network area on the basis of network environment analysis and adjusting transmission power level of network links are further research directions to enhance network capacity and reduce energy consumption.

\appendix
\begin{figure*}
\centering
\subfigure[]{\includegraphics[width=2.31in, trim=.8cm 0cm 1.3cm 0cm, clip=true]{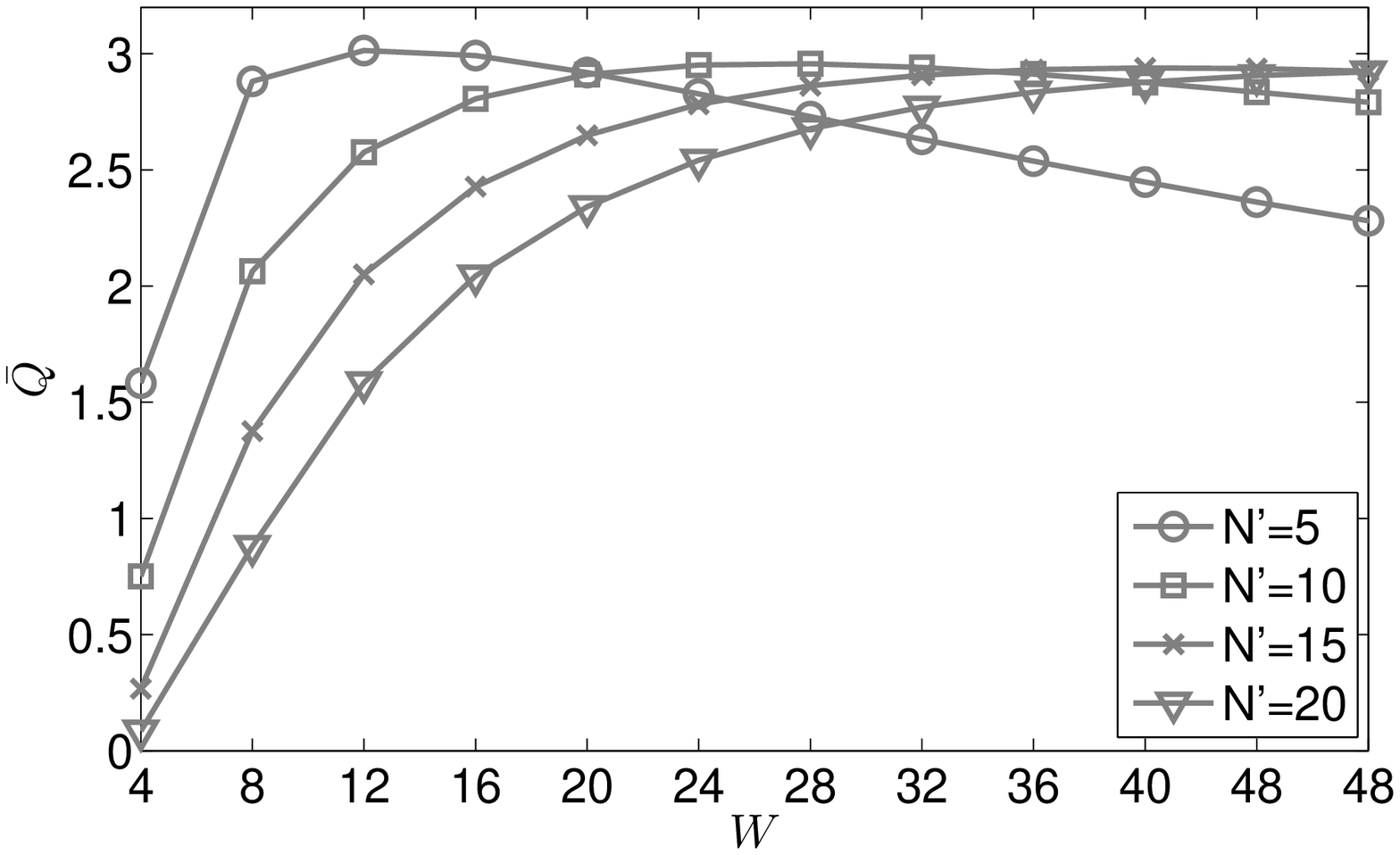}\label{NumberOfSuccess_CP}}
\subfigure[]{\includegraphics[width=2.31in, trim=.8cm 0cm 1.3cm 0cm, clip=true]{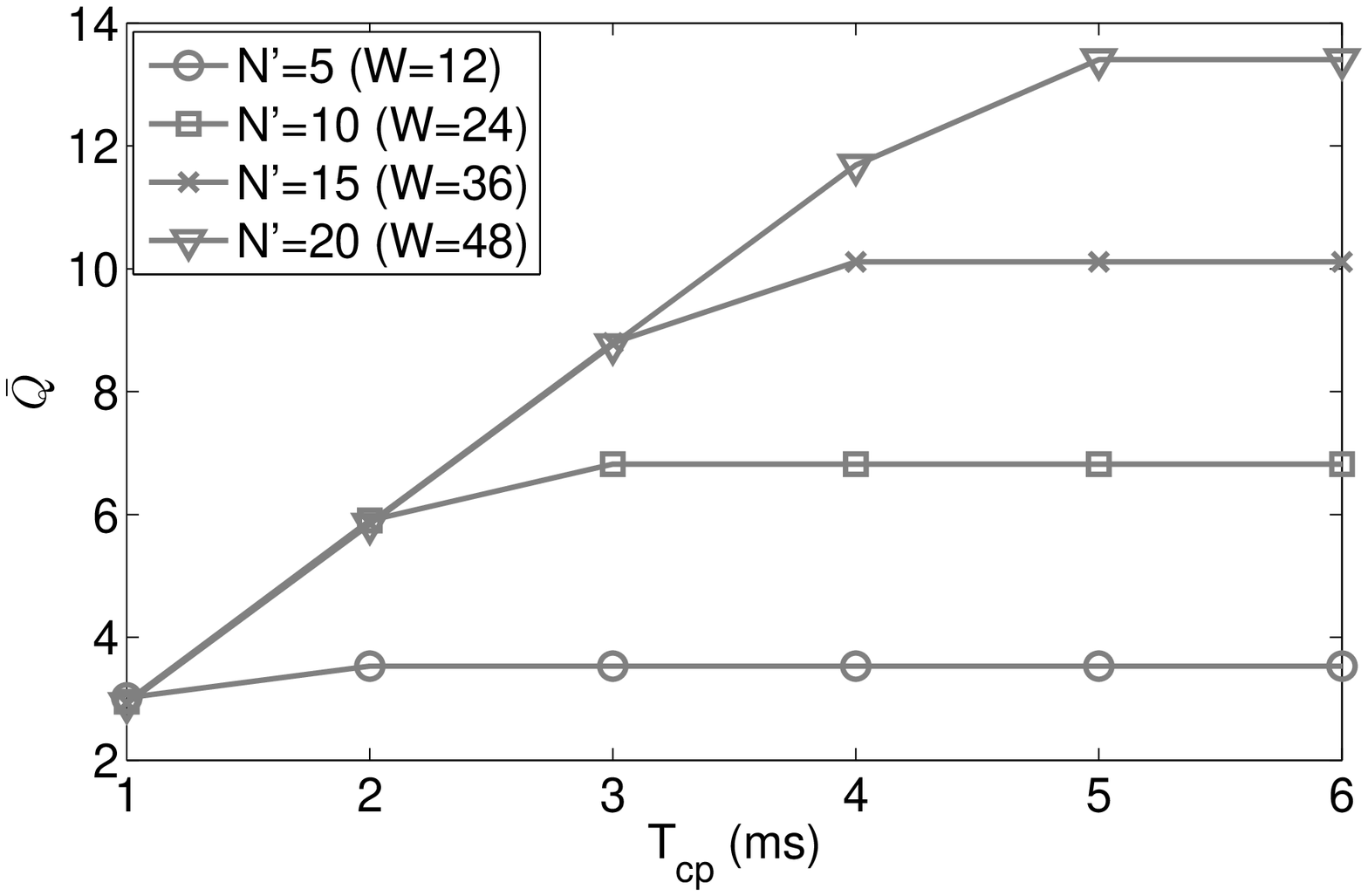}\label{NumberOfSuccess2_CP} }
\subfigure[]{\includegraphics[width=2.31in, trim=.8cm 0cm 1.3cm 0cm, clip=true]{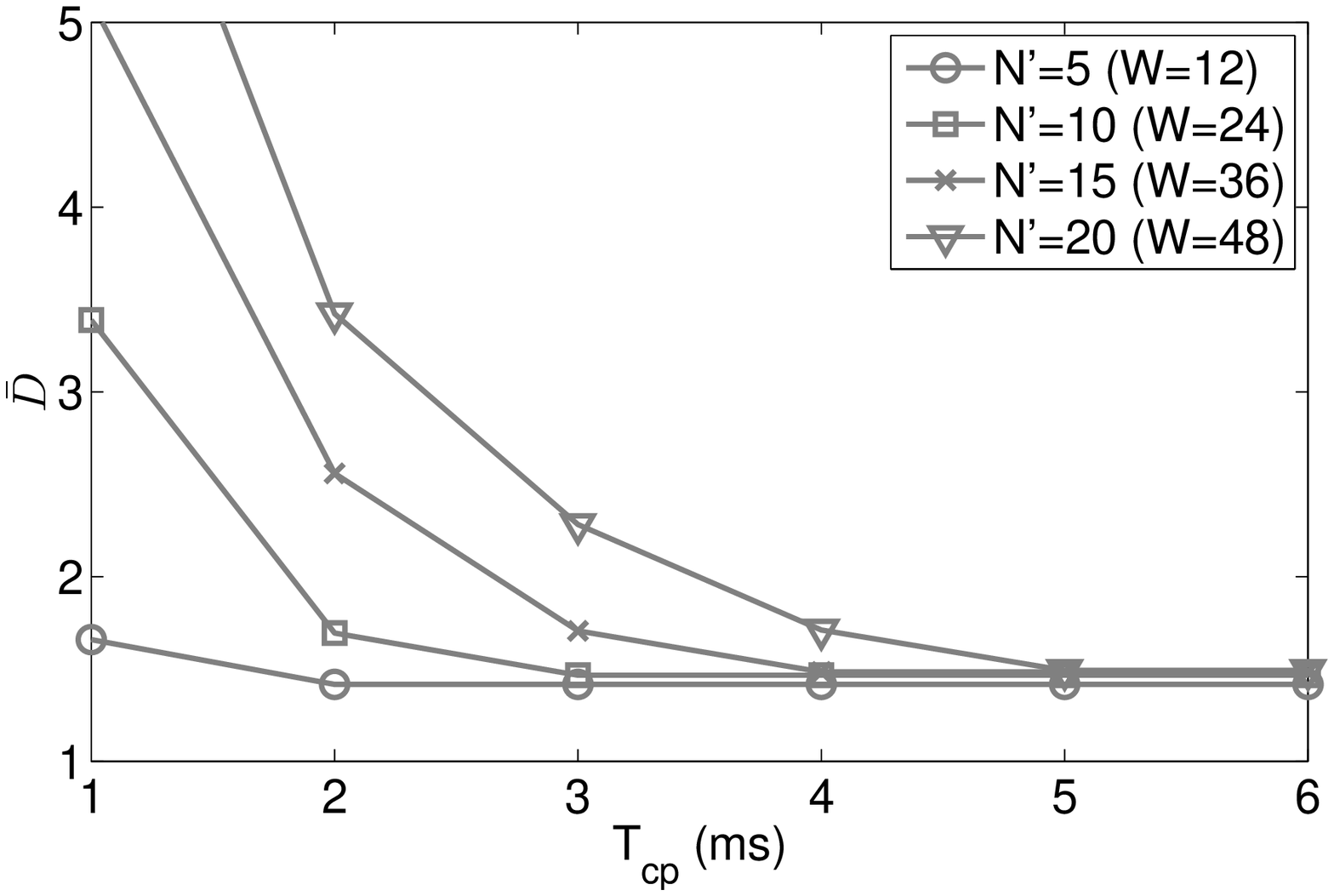} \label{Delay_CP}}
\caption{a) The number of successful transmission requests in 1 ms; b) The expected number of successful transmission requests in one frame (with the optimal contention window size); c) The average delay to initiate a new transmission normalized to frame duration (with the optimal contention window size).} \label{PerContention}
\end{figure*}

\section{Dynamic adjustment of the number of contention slots and contention window size}\label{Appen2}
In this section, we present a mathematical model to analyze the number of successful transmission requests in the contention slots and the average delay to initiate a new transmission. Based on the analytical model, we propose a mechanism to dynamically adjust the contention window size and the number of contention slots according to the traffic load and required delay to initiate a new transmission.

In the contention slots, the nodes that want to initiate a new transmission contend with each other using CSMA MAC to send a request packet to their cell coordinators. Each contending node chooses a random back-off time uniformly distributed in the range $[0,W-1]$, where $W$ is the contention window size that is dynamically set by coordinators. After each idle mini-slot, a contending node decreases its back-off window by one and transmits its request packet when its back-off window reaches zero. Nodes freeze their back-off window while the channel is busy and restart reducing the back-off window when the channel is idle again. We set the carrier sensing range, $r_c$, large enough (comparing to the maximum transmission range of requests, $r_g$) such that the hidden node problem is avoided, in order to reduce the probability of transmission request collisions. We also assume that contending nodes are uniformly distributed in the network area. Let $N'$ denote the number of contending nodes within a circular area with radius $r_c$. Thus, when a node starts to transmit a request packet, $N'-1$ other nodes (which are in the transmitting node's carrier sensing range) have to stay silent until the nodes finishes the transmission of its request packet. Let $T_{cp}$ denote the total duration of contention slots in a frame, $t_s$ denote the duration of a mini-slot, and $T_r$ denote the duration of a transmission request packet. Since contending nodes choose their back-off time uniformly distributed in the range $[0, W-1]$, when the channel is not busy a contending node starts to initiate transmission request packet in a mini-slot with probability $1/W$. Therefore, the probability that $X\in[0,N']$ nodes within a circular area with radius $r_c$ start transmission in a mini-slot (when the channel is not busy) can be written as
\begin{equation}\label{CP}
  p(X=i)= {N' \choose i} (\frac{1}{W})^{i} (1-\frac{1}{W})^{N'-i}.
\end{equation}
Using (\ref{CP}), the probability that a mini-slot is idle is
\begin{equation}\label{CP1}
  \delta_I=p(X=0)= (1-\frac{1}{W})^{N'},
\end{equation}
the probability of starting a successful transmission request in a mini-slot is
\begin{equation}\label{CP2}
  \delta_S=p(X=1)= \frac{N'}{W} (1-\frac{1}{W})^{N'-1},
\end{equation}
and the probability of a transmission collision in a mini-slot is
\begin{equation}\label{CP3}
  \delta_C=p(X\geq2)= 1-p_i-p_s.
\end{equation}
Consider a cycle as the time between two consecutive idle detection of mini-slots. The probability of initiating a transmission (successful or collision) after $M$ idle mini-slots is
\begin{equation}\label{CP4}
  P(M=m)=\delta_I^{m-1}(1-\delta_I).
\end{equation}
Thus, the average number of idle mini-slots in a cycle is
\begin{equation}\label{CP5}
  \bar{m}=\sum_{m\geq1}mP(m)=\frac{1}{1-\delta_I}=\frac{1}{1-(1-\frac{1}{W})^{N'}}
\end{equation}
and the average duration of a cycle is
\begin{equation}\label{CP6}
  \bar{T}_{cy}=\bar{m}t_s+T_r=\frac{t_s}{1-(1-\frac{1}{W})^{N'}}+T_r.
\end{equation}
Since the contention window size is $W$ and on average $\bar{m}$ idle mini-slots exits in a cycle, the expected number of cycles in the contention slots of a frame is
\begin{equation}\label{CP6b}
  \bar{u}=\min{(\frac{W}{\bar{m}},\frac{T_{cp}}{\bar{T}_{cy}})}=\min{(\frac{W}{\frac{1}{1-(1-\frac{1}{W})^{N'}}},\frac{T_{cp}}{\frac{t_s}{1-(1-\frac{1}{W})^{N'}}+T_r})}.
\end{equation}
Therefore, in a circular area with radius $r_c$, the expected number of successful transmission requests in the contention slots of one frame can be written as
\begin{multline}\label{CP7}
  \bar{Q}=\bar{u}\times\frac{\delta_s}{\delta_s+\delta_c}=\min{(\frac{W}{\frac{1}{1-(1-\frac{1}{W})^{N'}}},\frac{T_{cp}}{\frac{t_s}{1-(1-\frac{1}{W})^{N'}}+T_r})}\\ \times \frac{\frac{N'}{W} (1-\frac{1}{W})^{N'-1}}{1-(1-\frac{1}{W})^{N'}}.
\end{multline}
Figure \ref{NumberOfSuccess_CP} shows the the expected number of successful transmission requests (during 1 ms contention time) using different contention window sizes as the number of contending nodes varies. Figure \ref{NumberOfSuccess2_CP} shows the expected number of successful transmission requests in one frame as the total duration of contention slots increases (in each case, the contention window is optimized). Using (\ref{CP7}), the probability that a contending node successfully sends a transmission request to the coordinator in a frame is
\begin{equation}\label{CP8}
  \lambda_s=\frac{\bar{Q}}{N'}.
\end{equation}
Therefore, the probability that a node successfully sends its request to the coordinator after contending in $Y$ frames is
\begin{equation}\label{CP9}
  P(Y=y)=\lambda_s(1-\lambda_s)^{y-1}
\end{equation}
and the expected delay to initiate a new transmission is
\begin{equation}\label{CP10}
  \bar{D}=\sum_{y}yP(Y=y)T_f=\frac{T_f}{\lambda_s}=\frac{T_f N'}{\bar{Q}},
\end{equation}
where $T_f$ is the duration of a frame. Figure \ref{Delay_CP} shows the average delay to initiate a new transmission as the total contention slot duration increases for a different number of contending nodes in the carrier sensing range (the contention window is optimized).

The coordinators measure $\delta_I$, $\delta_S$ and $\delta_C$ by monitoring contention slots of the most recent frame. Based on (\ref{CP1}), (\ref{CP2}) and (\ref{CP3}) and the value of contention window size in the previous frame, they estimate the number of contending nodes within the carrier sensing range. The optimal value of contention window size can be calculated using (\ref{CP7}) for the estimated number of contending nodes. Also, the number of contention slots can be adjusted for the required transmission request delay using (\ref{CP10}). Accordingly, the number of contention slots and the contention window size are dynamically updated and announced by the coordinators.

\ifCLASSOPTIONcaptionsoff
\fi
\bibliographystyle{IEEEtran}
\bibliography{myref}
\vspace{-.5cm}

\begin{IEEEbiography}[{\includegraphics[width=1in,height=1.25in,clip,keepaspectratio]{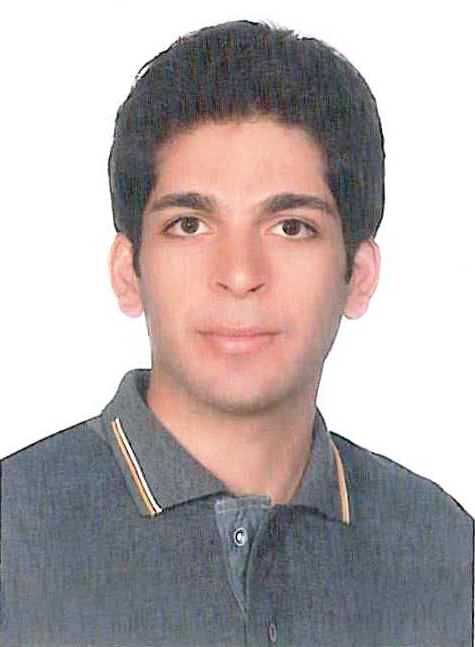}}]{Kamal Rahimi Malekshan} (S'11) is currently a PhD student at the Department of Electrical and Computer Engineering, University of Waterloo, ON, Canada. He has received M.Sc. degree in electrical engineering from the University of Tehran, Iran in 2011 and B.Sc. degree in electrical engineering from the University of Isfahan, Iran in 2008.

His current research interest include medium access control (MAC), power management and transmission power control (TPC) in wireless ad hoc networks.
\end{IEEEbiography}
\vspace{-.7cm}

\begin{IEEEbiography}[{\includegraphics[width=1in,height=1.25in,clip,keepaspectratio]{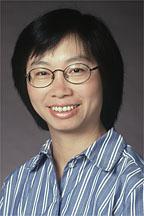}}]{Weihua Zhuang} (M'93-SM'01-F'08) has been with the Department of Electrical and Computer Engineering, University of Waterloo, Canada, since 1993, where she is a Professor and a Tier I Canada Research Chair in Wireless Communication Networks. Her current research focuses on resource allocation and QoS provisioning in wireless networks, and on smart grid. She is a co-recipient of several best paper awards from IEEE conferences. She received the Outstanding Performance Award 4 times since 2005 from the University of Waterloo and the Premier’s Research Excellence Award in 2001 from the Ontario Government. Dr. Zhuang was the Editor-in-Chief of IEEE Transactions on Vehicular Technology (2007-2013), and the Technical Program Symposia Chair of the IEEE Globecom 2011. She is a Fellow of the IEEE, a Fellow of the Canadian Academy of Engineering (CAE), a Fellow of the Engineering Institute of Canada (EIC), and an elected member in the Board of Governors and VP Mobile Radio of the IEEE Vehicular Technology Society. She was an IEEE Communications Society Distinguished Lecturer (2008-2011).
\end{IEEEbiography}

\vspace{-.7cm}
\begin{IEEEbiography}[{\includegraphics[width=1in,height=1.25in,clip,keepaspectratio]{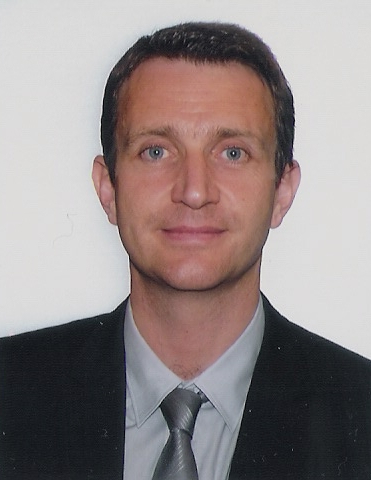}}]{Yves Lostanlen} (S'M98-M'01-SM'09) received the Dipl.-Ing (MSc EE) magna cum laude in 1996 from National Institute for Applied Sciences (INSA) in France. After three years of research at University College London and INSA he accomplished a European PhD summa cum laude in 2000. In 2009, he obtained a ScD (Habilitation) from University of Rennes, France and in 2013 he graduated from MIT Sloan School of Management, USA (Executive MBA.) Dr. Yves Lostanlen is currently CEO of SIRADEL North America and is based in Toronto, Canada, serving many top-tier companies in the ICT, Energy, Healthcare, Broadcast \& Media Industries: Government, policy makers, regulators, infrastructure operators and equipment manufacturers.
His current scientific interests are innovative technologies (hardware, software, data analytics) and services enabling energy-efficient infrastructure networks (telecom, energy, water) in under-developed regions in order to catalyze competitive advantage, productivity and growth.
Yves Lostanlen is also an Adjunct Professor in the Faculty of Applied Science and Engineering at the University of Toronto, Canada.
A senior member of IEEE, Yves Lostanlen has written more than 100 papers for international conferences, periodicals, book chapters and has been technical committee chairman, and session chairman at several international conferences. A frequent keynote speaker at international scientific and industrial conferences, Prof Lostanlen enjoys combining academic and industrial insights and technology and business constraints.
He received a "Young Scientist" Award" for two papers at the EuroEM 2000 conference.
\end{IEEEbiography}

\end{document}